\documentstyle[prl,aps,epsfig]{revtex}
\begin{document}


 \twocolumn[\hsize\textwidth\columnwidth\hsize  
 \csname @twocolumnfalse\endcsname              

\title{Universal features of the order-parameter 
fluctuations : reversible and irreversible aggregation}

\author
{Robert Botet$^{\dagger}$ and Marek P{\l}oszajczak$^{\ddagger}$}
 
\address{$^{\dagger}$
Laboratoire de Physique des Solides - CNRS, B\^{a}timent 510, Universit\'{e} Paris-Sud
\\ Centre d'Orsay, F-91405 Orsay, France  \\
and   \\  $^{\ddagger}$
Grand Acc\'{e}l\'{e}rateur National d'Ions Lourds (GANIL), \\
CEA/DSM -- CNRS/IN2P3, BP 5027,  F-14021 Caen Cedex, France  }
 
\date{\today}
 
\maketitle
 
\begin{abstract}
We discuss the universal scaling laws of order parameter
fluctuations in any system in which the second-order critical behaviour can be
identified. These scaling laws can be derived rigorously for equilibrium systems 
when combined with the finite-size scaling analysis. The relation between order
parameter, criticality and scaling law of fluctuations has been established
and the connexion between the scaling function and the critical exponents has 
been found. We give examples in out-of-equilibrium aggregation
models such as the Smoluchowski kinetic equations, or of at-equilibrium Ising 
and percolation models.
\end{abstract}


 ]  

\section{Introduction}
Fluctuations in many physical processes are difficult to analyze because they
develop dynamically and often keep the memory of initial conditions. 
On the other hand, strong fluctuations are ubiquitous, 
as show examples of hadronization in strong interaction physics, 
polymerization, colloid aggregation, aerosol coalescence or the formation of 
large scale structures in the Universe. With the advent of recently developed 
advanced detection
systems, the study of large fluctuations in physical observables 
became accessible in 'small systems', such as formed in
ultrarelativistic collisions of hadrons, leptons and nuclei, in the heavy-ion
collisions at the intermediate energies or in the collisions of atomic
aggregates \cite{leshouches}. 

In theoretical studies, it is
often assumed that fluctuations are irrelevant.
In this spirit, many aggregation processes have been
studied in the mean-field approximation \cite{kampenvan}. This problem has been
revisited recently \cite{aggreg}. It was shown
that contrary to the usual believe, fluctuations in
the size distribution  of largest cluster 
are generally large in the aggregation processes. Large fluctuations in the cluster
multiplicity distribution have been also reported in the binary fragmentation process
with the inactivation mechanism \cite{kno}.
 
This article deals with the features of fluctuations of physical quantities
in a $N$-body, $d$-dimensional system, with $N$ essentially {\it finite}. 
Moreover, the system is not necessarily at the thermodynamic equilibrium. Both these 
aspects of our approach are important in many areas of physics where, 
for example,  small and strongly fluctuating systems are produced 
in the violent collision processes. Consequently, these systems live
shortly and the typical time-scales are such that 
standard methods of equilibrium statistical physics 
may not to applicable.

We shall be particularly interested in self-similar systems, 
such as the fractal objects or thermodynamic systems at the second-order
phase transition. The self-similarity means in particular that one is unable 
to define the characteristic length $\sim (N^{*})^{1/d}$,
where $N^{*}$ is the characteristic size, which could be associated 
with the disappearance of fluctuations. Our aim in this work is the
discussion of universal scaling laws of fluctuations of different observables
in self-similar systems. In particular, we shall consider the order parameter
fluctuations in any system, both equilibrium and non-equilibrium one, in which the
second-order critical behaviour can be identified. These considerations will 
provide an understanding
of the relation between the order parameter, the criticality and the 
scaling law of fluctuations. The notion of {\it relevant variable} ({\it
relevant observable}) for the discussion of critical behaviour in finite 
systems will appear from this discussion.

The paper is organized as follows. In Sect. II the order parameter fluctuations
in statistical systems are analyzed and the relation with the
finite-size scaling analysis and Widom's hypothesis is developed in details
\cite{Widom} (Sect. II.B). The generic features of the tail of the scaling
function is addressed in Sect. II.C. In Sects. II.E - II.H, the generalized
scaling of the observable quantities , the $\Delta$-scaling, is discussed and
the reasons for the deviations from the limiting cases $\Delta=1$ and
$\Delta=1/2$ are presented.

Sects. III - VI are devoted to the detailed discussion of several well known generic 
models, using the results and methods of analysis proposed in Sect. II. In
Sect. III, the non-critical fragmentation model is discussed which exhibits the
power-law cluster size distribution. Results obtained in Potts model  are
discussed in Sect. IV. Sect. V is devoted to the discussion of the reversible
aggregation, as modelled by the percolation model. Both realistic 3-dimensional
bond percolation and the mean-field percolation on the Bethe lattice is being
considered. Irreversible aggregation and the fluctuation properties in the
Smoluchowski kinetic model are discussed in Sect. VI. Finally, the main
conclusions are given in Sect. VII.

\section{Order parameter fluctuations} 

\subsection{Correlation function argument}\label{sec:ornstein}

Let us call 
$m$ the observable under investigation. For the reason of presentation, 
we shall restrict ourselves to the case where $m$ is a scalar quantity 
and takes real positive values. ( In fact, this restriction is not a true 
limitation and, generally, one can consider $|m|^2$ as well.)    Fluctuations of the order parameter in thermodynamic systems 
are expected to have different properties at the critical point 
and outside of it. Far from the critical behaviour, the correlations
are short-ranged.
Fluctuations of the extensive order parameter $m$ in this case resemble the 
ergodic Brownian motion of this variable in its proper configuration space. 
Consequently ${<(m - <m>)^2>}/{<m>}$ is roughly a constant, meaning independent of 
the number of constituents in the sample. 
On the contrary, close to the second-order transition point, the 
fluctuations are correlated throughout the whole system and the correlation 
length $\xi$ for the infinite system becomes infinite as well.
Let us now define in such a system the deviation $\epsilon$ of 
the driven parameter from its critical value, $m$ the local value of the
order parameter and the field $h$ conjugated to $m$. The following isotropic 
correlation function is introduced :
\begin{eqnarray}
\label{def1}
\sigma_n(\epsilon,h,\vec{r}_1,...,\vec{r}_{n-1}) & = &
<m(\vec{r}_o) m(\vec{r}_o-\vec{r}_1)\dots \nonumber \\
&\dots& m(\vec{r}_o-\vec{r}_{n-1})> ~ \ .
\end{eqnarray}
Notation $<...>$ in the above expression denotes the thermodynamic 
average at a given $\epsilon$ and over the position $\vec{r}_o$. For vanishing 
$\epsilon$, all the length scales disappear and the correlation length $\xi$ 
diverges algebraically  as 
\begin{eqnarray}
\xi \sim \epsilon^{-\nu}  \nonumber 
\end{eqnarray}
with a universal exponent $\nu$, which depends only on the universality 
class of the transition. Moreover, the scaling description of second-order 
critical phenomena leads to the fundamental postulate that the thermodynamic 
potential $G$ verifies 
\begin{eqnarray}
\label{G}
G(\lambda \epsilon, \lambda^{2-\alpha -\beta} h) \sim 
\lambda^{2-\alpha} G(\epsilon,h) ~ \ ,
\end{eqnarray}
and this defines the two universal exponents $\alpha$ and $\beta$. Let us now 
go back to the correlation function $\sigma_n$ defined in (\ref{def1}). 
The integral of $\sigma_n$ over the $n-1$ space-variables : $\vec{r}_1,...,\vec{r}_{n-1}$, 
is equal to the $n$-th derivative of $G$ with respect to the field $h$. Hence, 
to be consistent with (\ref{G}), we must get the scaling relation :
\begin{eqnarray}
\label{sca1}
\sigma_n(\lambda \epsilon,\lambda^{2-\alpha -\beta} h,\lambda^{-\nu} 
\vec{r}_1,&\dots&, \lambda^{-\nu} \vec{r}_{n-1}) 
\sim \nonumber \\ &\sim&
\lambda^{n \beta} \sigma_n(\epsilon,h,\vec{r}_1,...,\vec{r}_{n-1}) ~ \ .
\end{eqnarray}
For $n=1$, we recover the well-known scaling behaviour of the
averaged order parameter with the critical exponent $\beta$. 
For any integer $n$, putting all the
space-variables to $\vec{0}$ in formula (\ref{sca1}) leads to the 
scaling of powers of the local order parameter :
\begin{eqnarray}
\label{pow}
<m^n>(\lambda \epsilon, \lambda^{2-\alpha -\beta} h) \sim \lambda^{n \beta}
<m^n>(\epsilon,h) ~ \ .
\end{eqnarray}
As a consequence, if we let $h=0$ and $\lambda=1/\epsilon$, one finds 
that the quantities $<m^n>/<m>^n$ are independent of $\epsilon$, when 
close to the transition. This means, by the finite-size analysis, that 
this ratio is independent of the size $N$ of the system at the transition point.

Let us now introduce the cumulants $\kappa_q$ from the general formula 
for the moment expansion of the order-parameter probability distribution $P[m]$ :
\begin{eqnarray}
\label{cum}
\ln \left( \sum_{m=0}^{\infty} P[m] \exp (m u) \right) = \sum_{q=0}^{\infty}
\frac{u^q}{q!} \kappa_q ~ \ .
\end{eqnarray}
Expanding the l.h.s. of the above expression in the power series in $u$, and
comparing corresponding powers on l.h.s. and r.h.s., one derives the 
relations between ordinary moments of $P[m]$ and the cumulant moments :
\begin{eqnarray}
\label{cum123}
\kappa_1 & = & <m> \nonumber \\
\kappa_2 & = & <m^2> - <m>^2 \nonumber \\
\kappa_3 & = & <m^3> - 3 <m^2><m> + 2 <m>^3  \\
\kappa_4 & = & <m^4> - 4 <m^3><m> - 3 <m^2>^2 + \nonumber \\
& + & 12 <m^2><m>^2 - 6 <m>^4 \nonumber \\
... \nonumber 
\end{eqnarray}
In the case of the second-order phase transition, as a result of scaling relations 
(\ref{pow}), all cumulant moments scale like  $$\kappa_q \sim <m>^q ~ \ .$$ 
Consequently, the generating function of the $m$-probability distribution (\ref{cum}) 
is only a  function of the reduced variable $<m> u$. This can be written as :
\begin{eqnarray}
\label{Pm}
P[m] = \frac{1}{2 \pi} \int_{0}^{2\pi} {\hat G}(i u) \exp(-im u) d u
\end{eqnarray}
where ${\hat G}$ is the generating function :
\begin{eqnarray}
{\hat G}(u) = \sum_{m=0}^{\infty} P[m] \exp(m u) \nonumber ~ \ . 
\end{eqnarray}
In the case when $<m>$ tends to $\infty$ but $m/<m>$ remains finite, 
we can rewrite formula (\ref{Pm}) in the more compact form :
\begin{eqnarray}
\label{first1}
<m>P[m] = \Phi\left( \frac{m}{<m>} \right) ~ \ ,
\end{eqnarray}
which is valid at the critical point.
$\Phi$ is the scaling function of the single reduced variable $m/<m>$.
As stated before, we can express this scaling as : 
\begin{eqnarray}
\label{iff}
\sum_{m=0}^{\infty} P_N[m] \exp (m u) = \Psi(<m> u)  ~ \ ,
\end{eqnarray}
which is the necessary and sufficient condition for the applicability of
scaling law (\ref{first1}) \cite{nowa}.
This result implies also that if this scaling occurs for 
the fluctuations of the parameter $m$ then it holds also for fluctuations 
of any power : $$X = N^a m^b$$ of this parameter as well. This is a consequence 
of (\ref{pow}) and $$P_N[X]dX = P_N[m]dm  ~ \ . $$

Till now, we did not specify reasons of changing $<m>$. Indeed, 
under the condition that the scaling framework of second-order phase transition 
holds, the scaling relation (\ref{first1}) is valid independently of the
explicit
reasons of changing $<m>$, and independently of any phenomenological 
details. In other words, the explicit relation between  the size $N$ of the
system and $<m>$ need not to be known at this stage. In the following section, 
we shall show how to derive supplementary informations about  $\Phi$  when the system
is at the pseudo-critical point.

\subsection{Widom's hypothesis and the finite-size scaling argument}
The hypothesis of Widom states that in the thermodynamic limit of
a system at thermal equilibrium, the free energy 
density close to the critical point scales as \cite{Widom} :
\begin{eqnarray}
f (\lambda^{\beta} \eta, \lambda \epsilon) \sim 
\lambda^{2-\alpha} f (\eta, \epsilon) ~ \ ,
\end{eqnarray}
where $\alpha, \beta $ are the usual critical exponents, $\eta$ is 
the intensive order parameter and $\lambda$ is the scale parameter. 
Even though finite systems do not exhibit the critical behaviour, 
nevertheless their properties may resemble those of infinite systems 
if the correlation length $\xi $ is larger or comparable to the typical 
length $L$~ of the system. In this case, one speaks about the pseudocritical 
point in a finite system at a distance  
\begin{eqnarray}
\label{row1}
\epsilon \sim c N^{-1/{\nu d}} 
\end{eqnarray}
from the true critical point \cite{FF}. The quantity $N$ in (\ref{row1}) is the size of the
$d$-dimensional system and $c$~ is some dimensionless constant which can be either positive 
or negative. This constant $c$ is negative if the maximum of finite-size 
susceptibility or of any other divergent macroscopic quantity in the thermodynamic 
limit lies in the ordered phase, while $c$ is positive if this maximum is in the disordered 
phase. One can then derive the finite-size scaling of the total free energy :
$$F(\eta , \epsilon, N) = N f (\eta , \epsilon )$$ at the pseudocritical point :
\begin{eqnarray}
\label{Widomfinite}
F_c(\eta ,N) \sim f(\eta N^{\frac{\beta }{2-\alpha}}) ~ \ .
\end{eqnarray}
In deriving (\ref{Widomfinite}) we used the hyperscaling relation : 
$$2-\alpha =  \nu d ~ \ . $$ The canonical probability density of the order 
parameter $P_N[\eta ]$ is given by \cite{Mayer} :
\begin{eqnarray}
\label{psi}
P_N[\eta ] = \frac{1}{Z_N} \exp(-{\beta}_T F(\eta , \epsilon , N)) ~ \ ,
\end{eqnarray}
where the coefficient ${\beta}_T ( \equiv 1/T )$ is independent of $\eta $ 
($T$ is the temperature of the system). Using Eq.(\ref{psi}), one may calculate 
the most probable value of the order parameter, which is the solution to the equation  
\begin{eqnarray}
\frac{\partial P_N[\eta]}{ \partial \eta} = 0 ~ \ , \nonumber  
\end{eqnarray}
as well as the average value of the order parameter and the partition function 
\begin{eqnarray}
\label{part}
Z_N \sim N^{-\frac{\beta }{2-\alpha}} \sim ~<\eta >~ \sim ~\eta^{*} ~ \ .
\end{eqnarray}
${\eta}^{*}$ in (\ref{part}) denotes the most probable value of the order
parameter. The average value of the order parameter vanishes for large values of
$N$, since both $\beta$ and $2-\alpha = 2 \beta + \gamma$ are positive.
The probability density $P_N[\eta ]$~ obeys then the scaling law
formally identical to Eq. (\ref{first1}) :
\begin{eqnarray}
\label{first}
<\eta > P_N[\eta ] = \Phi \left( \frac{\eta}{< \eta >} \right) \equiv \Phi (z) 
\end{eqnarray}
where, in addition,
\begin{eqnarray}
\label{firstsc}
\Phi (z) \sim \exp \left( - {\beta}_T f \left(a z, c \right) \right)  ~ \ .
\end{eqnarray}
In the above formula, we have omitted the temperature-dependent multiplicative 
factor which can be determined by normalization of $P_N[\eta ]$. 
Coefficients $a$ and $c $ may both depend on $\beta_T$.
We can then rewrite the scaling 
(\ref{first}) in a standard form for the extensive order parameter 
$m = N \eta$ :
\begin{eqnarray}
\label{stand1}
<m> P_N[m] = \Phi(z_{(1)}) 
\end{eqnarray}    
with the scaling variable $z_{(1)}$ defined by 
\begin{eqnarray}
\label{var1}
z_{(1)} = \frac{m-m^{*}}{<m>} ~ \ .
\end{eqnarray}
$m^{*}$ denotes the most probable value of the extensive order parameter. 
We call (\ref{stand1}) with (\ref{var1}) {\it the first-scaling law}.
The scaling domain is defined by this asymptotic behaviour of $P_N[m]$
when $m \rightarrow \infty$ and $<m> \rightarrow \infty$, but $z_{(1)}$ 
has a finite value. The normalization of the probability distribution 
$P_N[\eta ]$ and the definition of the average value of $m $, provide the two 
constraints :
\begin{eqnarray}
\label{eq17}
\lim \int_{-m^{*}/<m>}^{\infty} \Phi (z_{(1)}) dz_{(1)} & = & 1  \nonumber \\  & &
\nonumber \\
\lim \int_{-m^{*}/<m>}^{\infty} z_{(1)} \Phi (z_{(1)}) dz_{(1)} & = & 0~~~ \ \nonumber 
\end{eqnarray}       
The first-scaling law (\ref{first}) is a consequence of the self-similarity 
of the statistical system. The self-similarity means here that the 
fluctuations of the {\it reduced} order parameter 
$\eta /<\eta>$ at different 
scales characterized by different values of the intensive order parameter 
$<\eta>$,  have {\it identical properties}. This is a qualitative explanation 
for this scaling.

The logarithm of scaling function (\ref{firstsc}) corresponds to the 
non-critical free energy density at the renormalized distance $\epsilon =c$~ 
from the critical point. If it happens that the order parameter is related to 
the number of fragments, like in the Fragmentation - Inactivation - Binary (FIB) 
process \cite{sing1,sing11}, then (\ref{first}) can be written in an 
equivalent form to the Koba - Nielsen - Olesen (KNO) scaling \cite{koba}~,
proposed some time ago as an ultimate symmetry of the $S$ - matrix in 
the relativistic field theory\cite{comment1}~. The multiplicity distribution of
produced particles is intensely studied in the strong interaction physics where 
simple behaviour of much of the data on hadron-multiplicity distribution seems 
to point to some universality independent of the particular dynamical process. 

If, instead of real positive scalar, the parameter under investigation is a 
vector of dimensionality $n$ : 
$\vec {m} = [m_1, \cdots , m_n]$ , then the first-scaling law (\ref{stand1}, 
\ref{var1}) takes the more general form :
\begin{eqnarray}
\label{ndims1}
<|\vec{m}|>^n P_N[\vec{m}] = \Phi(\vec{z}_{(1)}) 
\end{eqnarray}
with 
\begin{eqnarray}
\label{ndimv1}
\vec{z}_{(1)} = \frac{\vec{m}-\vec{m}^{*}}{<|\vec{m}|>} ~ \ .
\end{eqnarray}
The scaling limit in (\ref{ndims1}) is defined by the asymptotic behaviour of 
$P_N[\vec{m}]$ when $m_i \rightarrow \infty$ ($i = 1, \cdots ,N$) and 
$<|\vec{m}|> ~\rightarrow \infty$, but ${z}_{i (1)}$~ ($i = 1, \cdots ,N$) have 
finite values.

\subsection{The tail of the scaling function}

The scaling function $\Phi$ introduced in the first-scaling law (\ref{stand1})
has some typical features reminiscent of the non-Gaussian critical distribution
of the order parameter. In this section, we are interested in the behaviour of 
scaling function for large values of the reduced parameter $m/<m>$, so we have
to study the system subject to the small field $h$ conjugated to the order
parameter. This breaks the symmetry of the distribution by shifting $m$ and $<m>$
towards larger values. More precisely, let us write the probability to get
the value $\eta$ of intensive order parameter at the distance $\epsilon$ from the 
critical point as :
\begin{eqnarray}
\label{Ph}
P_N[\eta,\epsilon,h] = P_N[\eta,\epsilon,0] 
\exp ({\beta_T \eta Nh}) ~ \ .
\end{eqnarray}
Till now, we have studied the behaviour of $P_N[\eta,\epsilon,0]$ for which the
first-scaling law holds when $\epsilon =0$ (the critical point) or 
$\epsilon = c N^{-1/\nu d}$ (the pseudo-critical point). 
Substituting (\ref{first}) and using  (\ref{part}), we obtain :
\begin{eqnarray}
\label{Phscal}
P_N[\eta,0,h] = N^{\frac{\beta}{2-\alpha}} \exp \left( 
{\ln \Phi \left( \eta N^{\frac{\beta}{2-\alpha}} \right) + 
\beta_T \eta Nh} \right) ~ \ .  \nonumber 
\end{eqnarray}
The most probable value $$\eta^{*} \sim h^{1/\delta}$$ of the order parameter 
in the limit of small external field $h$, is given by the maximum of the term 
in the exponential. Since $$\delta = \frac{2-\alpha-\beta}{\beta}$$  we get :
\begin{eqnarray}
\label{maxf}
\ln \Phi(h^{\frac{1}{\delta}} N^{\frac{1}{\delta+1}}) 
\sim -\beta_T h^{\frac{1}{\delta}}Nh ~ \ .
\end{eqnarray}
Relation (\ref{maxf}) is valid for any value of $N$ if and only if 
\begin{eqnarray}
\label{tail}
\Phi(z) \sim \exp ({-a z^{\delta + 1}}) \equiv \exp ({-a z^{\hat \nu}}) ~ \ ,
\end{eqnarray}
with the coefficient $a$ which depends on the temperature regularly.

One can express this relation in a different way. The anomalous dimension for  
extensive quantity $m = N \eta$ can be defined as : 
\begin{eqnarray}
\label{anomdim}
g = \lim_{N \rightarrow \infty} g_N = 
\lim_{N \rightarrow \infty} \frac{d}{d\ln N} \left( \ln <m> \right) ~ \ . 
\end{eqnarray}
One can see that due to both (\ref{part}) and the Rushbrooke relation between critical 
exponents :
\begin{eqnarray}
\alpha+2 \beta+ \gamma = 2 ~ \ , \nonumber  
\end{eqnarray}
the anomalous dimension is :  
\begin{eqnarray}
\label{an1}
g=1- \frac{\beta }{\gamma + 2 \beta } ~ \ . 
\end{eqnarray}
Since both $\alpha$ and $\beta$ are positive, therefore 
$g$ is contained between 1/2 and 1 for equilibrium systems at the critical point of 
the second-order phase transition . Because of these 
additional relations between critical exponents, one may note that a behaviour of the 
tail of the scaling function (\ref{tail}) is governed by the exponent :
\begin{eqnarray}
\label{gdab}
{\hat \nu} = \frac{1}{1-g} = \delta +1 = \frac{2-\alpha}{\beta}  ~ \ ,
\end{eqnarray}
which is always larger than 2. The limiting case : ${\hat \nu} = 2$ , 
{\it i.e.} the Gaussian tail (see 
(\ref{tail})), corresponds in this framework to the non-critical system. 

Finally, let us mention in passing that whenever cluster-size can be defined 
in a system exhibiting the second-order phase transition , like {\it e.g.} 
in the case of percolation, 
Ising model or Fisher droplet model, the exponent $\tau$ of the power-law cluster-size 
distribution  
\begin{eqnarray}
\label{sized}
n_s \sim s^{-\tau }   \nonumber 
\end{eqnarray} 
satisfies additional relations \cite{Stauffer} : 
\begin{eqnarray}
\gamma + \beta & = & \frac{1}{\sigma}   \nonumber \\ 
\nonumber \\
\gamma + 2\beta & = & \frac{\tau -1}{\sigma} \nonumber
\end{eqnarray}
what yields : 
\begin{eqnarray}\
\label{g}
g = \frac{1}{\tau -1} 
\end{eqnarray}
and
\begin{eqnarray}
\label{gdab1}
{\hat \nu} = \frac{\tau -1}{\tau -2}
\end{eqnarray}
Since $g$ at the second-order equilibrium phase transition is contained 
between 1/2 and 1, therefore the allowed values of exponent $\tau$ at the
critical point are : $2 < \tau < 3$, 
and the normalized cluster-size distribution : $$\sum_{s=1}^{N} s n_s = N$$
is $$n_s = N s^{-\tau} ~ \ . $$ 
\noindent
Consequently, whenever $\tau$ is defined, we get an information  
whether the studied equilibrium system is at the
second-order phase transition
and whether the considered extensive quantity can be identified with 
the order parameter of this transition.

Let us define, for example, the multiplicity as the total number of 
clusters. (The definition of clusters includes also the monomers.) 
The cluster multiplicity cannot be an 
order parameter of these kind of equilibrium phase transitions
because with  $2 < \tau < 3$ :
\begin{eqnarray}
\sum_{s=1}^{N} n_s \sim N ~ \ , \nonumber
\end{eqnarray}
what means that the average multiplicity scales as 
the total mass of the system at the transition point, {\it i.e.} $g=1$. 
On the other hand, the size of the largest cluster is a natural order 
parameter for these kind of phase transitions. In this case 
we have : 
\begin{eqnarray}
\label{eq28}
<s_{max}> \sim N^{\frac{1}{\tau -1}}  ~ \ ,
\end{eqnarray}
what is a direct consequence of : $$ \sum_{s=<s_{max}>}^{N} n_s \sim 1 ~ \ , $$ {\it
i.e.} that there is in the average only one largest cluster. 
Moreover, the relation (\ref{g}) derived for the 
second-order critical phenomenon is correctly recovered. 
One should emphasize, that the relation 
(\ref{eq28}) is very general and its derivation does not depend 
on the assumption of thermodynamic equilibrium. In other words, the
relation (\ref{eq28}) between the anomalous dimension and the exponent $\tau$ 
, is valid also for the off-equilibrium second-order 
phase transitions. We shall return to this point in Sect. VI.

The cluster multiplicity
could be the order parameter whenever $\tau < 2$, though this cannot happen in
the equilibrium phase transitions. Note that this argument to sort among 
different  candidates for the
order parameter , requires only the knowledge of $\tau$, {\it i.e.}
the complete information about the critical process is
superfluous. We shall use this argument later  in the case of
percolation model and Smoluchowski model of gelation.
Finally, we shall see below in the Mekjian model 
that we may have power-law size distribution 
with $\tau < 2$, in the absence of phase transition governed by the  
multiplicity as the order parameter.

\subsection{Landau - Ginzburg theory of phase transitions}

Let us consider the Landau - Ginzburg (LG) theory as an exactly 
solvable example of the second-order phase transition. 
The homogeneous LG free energy density is :
\begin{eqnarray}
\label{free}
f(\eta ) = \epsilon \eta^2 + b \eta^4 + \cdots \nonumber
\end{eqnarray}
where $b$ is a positive constant. The most probable value of the order parameter 
$\eta $ in the disordered phase ($\epsilon < 0$) is implicitly set to 0. 
It is more convenient to work with the extensive order parameter  $m = N \eta$ when 
dealing with the finite systems.
The probability of a state $m$ for a given $\epsilon$  is \cite{Mayer} :
\begin{eqnarray}
\label{proba}
P_N[m] =\frac{1}{Z_N} \exp \left[ -\beta_T(\epsilon \frac{m^2}{N} + 
b \frac{m^4}{N^3} - \cdots) \right] ~ \ .
\end{eqnarray}
$Z_N$ is defined by the normalization of $P_N[m]$. 
To remain consistent with other sections of this paper and without loss 
of generality, we consider now the case where $m$ is positive.
We will admit that $N$ is so large that the first two terms in the free energy 
expansion are sufficient to study the phase transition.
At the critical point : $\epsilon =0$, the leading term of the free energy 
density is proportional to $m^4$. Standard integrations yield the values for
the partition function $Z_N$ and the average value of the order parameter
$<m>$, both proportional to $N^{3/4}$. 
Introducing them in (\ref{proba}), one finds :
\begin{eqnarray}
\label{firstLG}
<m>P_N[m] & = & \frac{4 \sqrt{\pi}}{\Gamma^2[1/4] } \times \nonumber \\&\times&
\exp \left( -\frac {\pi^2}{\Gamma^4[1/4]}
\left( \frac{m}{<m>} \right)^{4} \right)  ~ \ ,
\end{eqnarray}
which has the form of (\ref{first1}). Note that the scaling function : 
$\Phi(z) \sim \exp (-z^4)$, decreases very fast as one moves away from the most 
probable value. This result is consistent with the analysis done in the previous section.

The pseudo-critical point is the value of $\epsilon$ for which the finite-size 
thermal susceptibility reaches its maximum. Writing that the inverse of this
susceptibility is the second derivative of the free energy with respect to the order 
parameter, one finds : 
\begin{eqnarray}
\label{epsilon}
\epsilon = - 6\frac{\Gamma[3/4]}{\Gamma[1/4]} \left( \frac{b}{\beta_T N}
\right)^{1/2} ~ \ . 
\end{eqnarray} 
This results is correct at the 
first order in $N^{-1/2}$. Replacing $\epsilon$ in (\ref{proba}) by
(\ref{epsilon}), leads to the scaling form of $P_{N}[m]$ :

\begin{eqnarray}
<m> P_N[m] & = & A \exp \left[ - \frac{\Gamma[3/4]^2}{\Gamma[1/4]^2}
((m/<m>)^4 \right. -\nonumber \\& - & \left. 6 (m/<m>)^2) \right] ~ \ ,
\end{eqnarray}
where $A$ denotes a normalization constant. We recover indeed the first-scaling law with the 
exponential tail : $\exp (-a z^4)$, for the large arguments.

Outside of the critical point in the disordered phase ($\epsilon >0$), the leading term 
of the free energy is proportional to $m^2$, and the probability distribution 
$P_N[m]$ is essentially Gaussian. Deriving, as previously, the values of $Z_N$ and 
$<m>$ (both behave like $N^{1/2}$ in this case), we get the scaling form :
\begin{eqnarray}
\label{ht}
<m>P_N[m] = \frac{4}{\pi} \exp \left( -\frac{4}{\pi} \left( \frac{m}{<m>}
\right)^2 \right) 
\end{eqnarray}
which is still under the form (\ref{first}) but with a Gaussian scaling function
reminiscent of the Gaussian fluctuations.

Finally, in the low temperature regime ($\epsilon <0$), the most probable
value of the order parameter is positive : 
\begin{eqnarray}
m^{*}= \sqrt{-\frac{\epsilon}{2b}} N  ~ \ . \nonumber
\end{eqnarray}
Developing $P_N[m]$ in (\ref{proba}) around this point leads to the expression :
\begin{eqnarray}
\label{secondLG}
m^{* 1/2} P_N[m] &\simeq& \left(-2 \frac{\epsilon^3}{b \pi^2}
\right)^{1/4}\times \nonumber \\&\times&
\exp \left( \epsilon \sqrt{-2 \frac{\epsilon}{b}}~ \frac{(m-m^{*})^2}{m^{*}}  
\right)    ~ \ ,
\end{eqnarray}
which is no more in the standard form (\ref{first}). In this case, the average
value of the order parameter $<m>$ is of the same order of magnitude as its most 
probable value $m^{*}$ and one can rewrite (\ref{secondLG}) in the scaling form :
\begin{eqnarray}
\label{2LG2}                                                              
<m>^{1/2} P_N[m] \sim \exp \left( -a \frac{(m-m^{*})^2}{<m>} \right)  ~ \ ,
\end{eqnarray}
where $a$ is a positive constant. This particular scaling form will be
discussed later in details.

\subsection{The $\Delta$-scaling law}
One may ask, what happens if the observable 
quantity is not the order parameter 
but the $N$-dependent function of the order parameter like : 
\begin{eqnarray}
\label{delta0}  
m=N^{a_1 } - N^{a_2} \eta  ~ \ ,
\end{eqnarray}
where
\begin{eqnarray}
\label{acondition}
a_1 > g+a_2-1 \ . 
\end{eqnarray}
The latter condition ensures that the order parameter does not determine 
the leading behaviour of $m$. For large $N$ : $$<m>~ \sim N^{a_1}~.$$ 
Writing (\ref{stand1}) with $m$ instead of $\eta$ and taking into account that : 
$$P_N[\eta ] d\eta =P_N[m]dm$$ one finds the generalized law :
\begin{eqnarray}
\label{delta}
<m>^{\Delta}P_N[m] & = & \Phi (z_{({\Delta})}) 
\equiv \Phi \left( \frac{m-m^{*}}{<m>^{\Delta}} \right)   ~ \ ,
\end{eqnarray}
where :
\begin{eqnarray}
\label{deltawyr}
\Delta & = & \frac{g+a_2-1}{a_1} < 1  ~ \ . \nonumber  
\end{eqnarray}
This generalized law will be called in the following 
{\it the $\Delta$-scaling law}. The scaling 
function ${\Phi} (z_{(\Delta )})$ depends only on one scaled variable :
\begin{eqnarray}
\label{deltawar}
z_{(\Delta )} = \frac{{m-m^{*}}}{{<m>^{\Delta}}} ~ \ .
\end{eqnarray} 
The normalization of the probability distribution $P_N[m]$ and the definition of the 
average value of $m$ provide two constraints :
\begin{eqnarray}
\label{eq17x}
\lim \int_{-<m>^{1- \Delta }}^{\infty} \Phi (z_{\Delta}) dz_{\Delta} =1~~~  \nonumber \\
\nonumber \\
\lim \int_{-<m>^{1- \Delta }}^{\infty} z_{\Delta} \Phi (z_{\Delta}) dz_{\Delta} =0~~~  \nonumber 
\end{eqnarray}                                                
which are consistent with : $\Delta \le 1$, because the scaling function $\Phi $ is 
positive. The scaling function ${\Phi}(z_{(\Delta )})$ in (\ref{delta}) has an
identical form as ${\Phi}(z_{(1)})$, except for the inversion of the 
abscissa axis. 
In particular, its tail for large $z_{(\Delta )}$ has the same form :
\begin{eqnarray}
\label{deltatail}
\Phi (z_{(\Delta )}) \sim \exp \left( -z_{(\Delta )}^{{\hat \nu}} \right) = 
\exp \left( -z_{(\Delta )}^{\frac{1}{1-g}} \right)  
\end{eqnarray}
as given in (\ref{tail}). One should mention in passing that if $$a_1 < g + a_2-1$$ 
in (\ref{delta0}), then $$<m> \sim N^{a_2}<\eta >$$ and $\Delta = 1$,
following the remark of Sect. II.A.

According to (\ref{firstsc}), the logarithm of scaling function $\Phi (z_{(\Delta )})$ :  
\begin{eqnarray}
\label{log}
\ln \Phi (z_{(\Delta )}) = -{\beta}_T f(a z_{(\Delta )}, c)~ \  \nonumber 
\end{eqnarray} 
is related to the {\it non-critical} free energy $f$, in either ordered ($c>0$) or 
disordered $(c<0)$ phase.    

As an important example, we see from (\ref{delta0}), (\ref{delta}), that the 
$\Delta $-scaling of the extensive variable : 
\begin{eqnarray}
{\hat m} = N(1 - \eta ) \equiv N{\hat {\eta}} \nonumber 
\end{eqnarray} 
can be used to determine the anomalous dimension since in this case : $\Delta = g$. 
For this reason , ${\hat m}$ is a very useful variable in all phenomenological studies. 
At the phase transition : $$< N{\hat {\eta}} >~ \sim N$$ but the 
finite-size corrections are {\it algebraic}.

\subsection{Off-critical scaling}

$\Delta = 1/2$ with ${\Phi}(z_{({\Delta})})$ nearly Gaussian, is a particular case
of a $\Delta$-scaling associated with the non-critical systems 
\cite{prln}. This limit :
\begin{eqnarray}
\label{stand2}
<m>^{1/2} P_N[m] & = & \Phi \left( \frac{m-m^{*}}{<m>^{1/2}} \right) 
\nonumber \\ & \equiv & \Phi(z_{(1/2)})  ~ \ ,
\end{eqnarray}
which is called {\it the second-scaling law}, has been found in the shattering 
phase 
of the non-equilibrium FIB process \cite{nowa} and in the 'gaseous' phase of equilibrium 
percolation process \cite{prln}. 
We should recall also that this form of scaling function has been seen for LG model in the 
low temperature regime (see (\ref{2LG2})).

More generally, let us suppose now that the extensive parameter $m$ is not critical, 
{\it i.e.} either the system is in a critical state but the parameter 
$m$ is not critical, or the system is outside of the critical region. The 
value of $m$ at the equilibrium is obtained by minimizing the free 
energy. The free energy $F$ is analytical in the variable $m$ close to its 
most probable value $m^{*}$ :
\begin{eqnarray}
\label{freenot}
F \sim N^{-1} (m-m^{*})^{2} ~ \ .
\end{eqnarray}
Using (\ref{freenot}) one obtains $$<m> \sim \mu^{*} N~ , $$ where $\mu^{*}$ is a 
positive (finite) number independent of $N$ and : 
\begin{eqnarray}
Z_N \sim N^{1/2} \sim 
~<m>^{1/2} ~ \ .
\end{eqnarray}
The probability density $P_N[m]$ verifies the second-scaling law (\ref{delta}) :
\begin{eqnarray} 
\label{second}
<m>^{1/2} P_N[m] & = & \exp \left( -\beta_T \mu^{*} \left( 
\frac{m-<m>}{<m>^{1/2}} \right)^{2} \right)   \nonumber \\ 
& \equiv & \Phi (z_{(1/2)}) ~ \ .
\end{eqnarray} 
This is a particular case of the $\Delta$-scaling law ($\Delta=1/2$) and    
the scaling function is now Gaussian \cite{comment2}. This scaling 
(\ref{stand2}) holds for $<m> \sim N$ but now with the {\it exponential} finite-size 
corrections. This is a principle difference from the finite-size corrections and/or 
$\Delta$-scaling. The above arguments apply to any second order phase transition.
In particular, they are not limited to LG theory of phase transitions 
(see Eq. (\ref{2LG2})).

\subsection{Finite-size cross-over effects }
The discussion of previous section is valid for systems at the critical (and pseudo-critical) 
point or far from the critical point in the ordered phase. Let us suppose now that the
system is prepared such that : 
\begin{eqnarray} 
<m> \sim N^{g^{'}} ~~~~, ~~~~ g^{'} < 1  \nonumber
\end{eqnarray}     
and $g^{'}$ is not the anomalous exponent. Here, we would like to study how the finite 
system evolves when the control parameter $\epsilon$ tends slowly to 0, namely : 
\begin{eqnarray}
\epsilon \sim N^{2g^{'} -2} \nonumber ~ \ . 
\end{eqnarray}
We shall address this question in the mean-field approximation using LG theory. Let us 
first write down the average value :
\begin{eqnarray}
\label{mmoy}
<m> = \frac{\int_{0}^{\infty} m
\exp(-\epsilon m^2/N - b m^4/N^3) dm}
{\int_{0}^{\infty} \exp(-\epsilon m^2/N - b m^4/N^3) dm} ~ \ .
\end{eqnarray}
Hence, writing this definition with the new driving parameter
$\epsilon' = \epsilon N^{1/2}$ and using the rescaled variable $m'=m/N^{3/4}$, the average 
value of $m$ can be put in the form : 
\begin{eqnarray}
\label{fss}
<m> = N^{3/4} \psi(\epsilon N^{1/2}) ~ \ ,  \nonumber 
\end{eqnarray}
while its most probable value is : 
$$m^{*} = \sqrt{-\frac{\epsilon}{2b}}N~ .$$ If the exponent
$g^{'}$ is not too small, {\it i.e.} if $\epsilon$ does not vanish too fast, the two 
quantities : $<m>$ and $m^{*}$, have to coincide. This is because the exponential weight term 
in (\ref{mmoy}) diverges as $\sim \exp(\epsilon^2 N/4b)$ when  
\begin{eqnarray}
\epsilon N^{1/2} \sim N^{\frac{4g^{'}-3}{2}}  
\end{eqnarray} 
becomes large with increasing $N$. As a consequence,
the common behaviour of $<m>$ and $m^{*}$ is : $\sim N^{g^{'}}$. 
The scaling form (\ref{secondLG}) in this case is :
\begin{eqnarray}
\label{devia}
<m>^{\Delta} P_N[m] \sim \exp \left( -c \frac{(m-m^{*})^2}{<m>^{2\Delta}}
\right) 
\end{eqnarray}
with $c$ a positive constant, and : $$\Delta = \frac{3}{2}g^{'} - 1 ~ \ . $$
We recover here the two cases previously discussed in Sect. II.B. 
When $g^{'} = 1$, 
{\it i.e.} when $\epsilon = {\mbox {const}}$, then this is the second-scaling law. 
When $g^{'} = 3/4$, then this is the first-scaling law since the finite system is yet in 
the critical region ($g^{'} = g$). In between these two limiting cases, 
$\Delta$-scaling 
holds with $1/2 < \Delta < 1$. Note also that the scaling function in (\ref{devia}) has 
a Gaussian form, even for $\Delta > 1/2$, which is quite different from the case 
(\ref{tail}) of Sect. II.C.

\subsection{Summary : panorama of the $\Delta$-scaling for 
thermodynamic systems}
Several features of finite systems are important if one wants to study either the
criticality of the corresponding infinite system or the distance to the critical point.  
One should name here : the $\Delta$-scaling (this includes the first-scaling law 
$\Delta = 1$ as well), the form of the tail of the scaling function $\Phi$, and the 
anomalous exponent. 
All these features are closely related with the properties of the 
scaling function which characterizes finite system at the equilibrium. If the infinite 
system experiences a second-order phase transition, and if $m$ is the 
scalar order parameter or the shifted scalar order parameter 
(\ref{delta0}), then :
 
{\underline {At the critical point}}, the corresponding finite system exhibits 
first-scaling law if $m$ is the order parameter, or $\Delta$-scaling law if $m$ is the 
shifted order parameter. In both cases, the tail of the scaling function 
$\sim \exp(-z^{{\hat \nu}})$ is characterized by a large value of the exponent 
${\hat \nu} = 1/(1-g) >2$, with $g$ being the anomalous exponent, {\it i.e.} the exponent 
characterizing decrease of the extensive order parameter with the size $N$ of the finite 
system. The values of $\Delta$ are restricted to  $0 < \Delta \leq 1$, and the anomalous
exponent $g$ takes values in between 1/2 and 1 for second-order at-equilibrium 
phase transition. 

{\underline {Far from the critical point}}, finite system exhibits 
second-scaling law with the Gaussian tail of the scaling function. 

{\underline {Close to the critical point}} when $\epsilon \rightarrow 0$ if $N\rightarrow \infty$, 
the finite system exhibits the cross-over phenomenon from the first-scaling to the 
second-scaling law by a continuous $\Delta$-scaling law with Gaussian shape of the 
scaling 
function. One should remind here that the precise dependence of $\epsilon (N)$ is 
irrelevant provided that : $g < g^{'} <1$, {\it i.e.} that the conditional point 
approaches 0 not faster than the pseudocritical point. This last remark is important in 
the phenomenological applications of the scaling theory to the situations where the 
$N$ - dependence of the conditional point is governed by an external control parameter with
unknown relation to the system size. 

Last but not least, if the parameter $m$ is not 
singular at the transition, then all properties of its probability distribution are the 
same as in the case of non-critical systems.

In phenomenological applications, it is often difficult to get the probability 
distribution with sufficient accuracy for the values of scaling variable which are far 
from the most probable value since this corresponds to very small probabilities. It is 
then more judicious to work with moments of the distribution instead of the distribution 
itself. For example, when the system undergoes $\Delta$-scaling, the properly
normalized cumulant moments (\ref{cum123}) :
\begin{eqnarray}
\label{scamom}
\frac{\kappa_q}{(\kappa_1)^{q \Delta }} \sim {\mbox {const}} ~ \ ,
\end{eqnarray}
are independent of the size of the system. An important consequence
is that the generating function of the $m$-distribution :
${\hat G}(u) = \sum P[m] \exp(m u)$, is a function of the 
reduced variable $<m>^{\Delta} u$ only, generalizing a remark of Sect. II.A
for the generating function in the first-scaling case.

\section{A non-critical model : the Mekjian model}
The Mekjian fragmentation model is an equilibrium model which describes the decomposition 
of system into an ensemble of fragments. The statistical weights for every configuration 
of fragments are given explicitly in this model. If $n_s$ denotes the number of fragments 
of size $s$ with the size-conservation : $N=\sum_s s n_s$, the weight function for the 
configuration $\{n_s\}$ is given by \cite{Mek1} :
\begin{eqnarray}
\label{Mekjian}
W_N(\{n_s\},x) = \prod_{s=1}^{N} \frac {s x^{n_s}}{n_s! s^{n_s} (x+s-1)} 
~ \ ,  \nonumber  
\end{eqnarray}
with $x$ being a real control parameter. Many exact results can be obtained in this simple 
model. Here, we are interested in the multiplicity distribution $P_N[m]$, where the 
fragment multiplicity is : $m=\sum_s n_s$. We can show that \cite{Mek2} :
\begin{eqnarray}
\label{PNM}
P_N[m] = x^m |S_N^{(m)}| \frac {\Gamma(x)}{\Gamma(N+x)} ~ \ ,  \nonumber 
\end{eqnarray}
where $|S_N^{(m)}|$ are the signless Stirling numbers of the first kind. 
Knowing then the generating function for these Stirling numbers :
\begin{eqnarray}
\label{gfstir}
\sum_{m=0}^{\infty} P_N[m] e^{m u} = \frac{\Gamma(x) \Gamma(xe^{u}+N)}
{\Gamma(xe^{u}) \Gamma(x+N)} ~ \,
\end{eqnarray}
one obtains the average value of $m$ :
\begin{eqnarray}
\label{averm}
<m> & = & x \sum_{s=1}^N \frac{1}{x+s}\nonumber \\& = & 
x \ln N +(x-1)\gamma-\psi(x) + {\cal O} (1/N)  \nonumber ~ \ .
\end{eqnarray}
Moreover, making an asymptotic development of  (\ref{gfstir}) for large $N$ and small $s$, 
one obtains :
\begin{eqnarray}
\label{Poisson}
\sum_{m=0}^{\infty} P_N[m] e^{m u} \simeq N^{x(e^{u}-1)} ~ \ .
\end{eqnarray}
The latter approximation is known to be correct for finite values of $u$ \cite{Delannay}. This means 
that $P_N[m]$ is approximately a Poisson $m$-distribution with parameter $x \ln N$. 
In the leading order we have then : $<m>~ \simeq ~m^{*}$ . 
Inverting (\ref{gfstir}) to get $P_N[m]$ as a Fourier transform, and making $N$ large 
yields the scaling formula :
\begin{eqnarray}
\label{Msca2}
<m>^{1/2} P_N[m] & = & \frac{1}{\sqrt{ \pi}} \exp 
\left( -\frac{(m-<m>)^2}{2<m>}- \right.\nonumber \\ & - & (x-1)(\gamma-\psi(x))
\frac{m-<m>}{<m>} + \nonumber \\ & + & \left. 
{\cal {O}}\left( \frac{1}{<m>} \right) \right) ~  \ .
\end{eqnarray}
This is nothing else but the second-scaling law (\ref{stand2}) for the multiplicity 
distribution when $N$ becomes large enough, because $<m> \simeq m^{*}$ . 
When $<m>$ is large enough, the second term in (\ref{Msca2}) is always very small compared 
to the first one for a finite $x$. 

Different fixed values of the control parameter $x$ mimic different situations of the 
fragmentation. For $x \ll 1$, one has the situation of a fused system. For $x \sim 0.5$, 
the fragmentation resembles the evaporation of light fragments. 
The limit $x \gg 1$ corresponds to the complete dissociation of the mass into light 
fragments (monomers). Each of this situation is characterized by a different 
fragment-size distribution. The case $x=1$ is particular in this model since it leads to
the power-law size-distribution (\ref{sized}) with the exponent $\tau = 1$.
Following discussion in Sect. II.C, cluster multiplicity could be the order
parameter. On the other hand, the second-order equilibrium phase transition is
associated with $2 < \tau < 3$, what implies that the equilibrium model of  
Mekjian is a non-critical model. Indeed, that is what can be seen also in the 
cluster-multiplicity scaling law (\ref{Msca2}).
 Hence, the power-law cluster-size distribution alone does not
guarantee that the system exhibits the critical behaviour of any kind 
\cite{sing11}.

\section{Example : The Potts model}

A generalization of the magnetic Ising spin model has been proposed by Domb 
\cite{Domb} and studied in details by Potts \cite{Potts}. In this model, one
considers a system of $N$ sites in the $d$-dimensional space. The magnetic state 
of each site $i$ is characterized by a quantity called : a spin (say : $s_i$). 
Each spin is of the same constant modulus and points to one of the 
$q$ equally spaced 
directions, labelled from 0 to $q-1$. The ferromagnetic short-ranged Potts 
Hamiltonian is then :
\begin{eqnarray}
H_q = -J \sum_{i,j} \delta (s_i,s_j)     ~ \ ,
\end{eqnarray}
where $\delta$ is the Kronecker symbol, and $J$ is the positive coupling constant.
The sum is restricted to nearest-neighbour pairs. 
The site percolation corresponds to
the $q=1$ Potts model, and ferromagnetic Ising model to the $q=2$ case.
This model is one of the simplest non-trivial 
critical thermodynamic N-body system,
and many exact or accurate results are known for standard values of $(d,q)$. 
In particular, there exists a value $q_c(d)$ 
(for example : $q_c(2)=4$), for which
when $q \le q_c(d)$, then such an interacting system 
experiences a second-order phase 
transition at a finite critical temperature (for example : 
$\beta_c J= \ln(1+ \sqrt{q})$ at $d=2$), 
while for $q>q_c(d)$ the transition is a first-order one.

Let us consider here the case of the second order phase transition. All the 
scalings
described above should hold. We have first to define the order parameter for the 
system. If for a given configuration of the system, we call $N_k$ the number of 
sites in the state $k$, where $k$ varies from 0 to $q-1$, then 
the order parameter $m$ is given by :
\begin{eqnarray}
\label{eq53}
m=\frac{q(N_{max}/N)-1}{q-1}                 ~ \ .
\end{eqnarray}
$N_{max}$ in (\ref{eq53}) is defined as the maximum of all $N_k$'s.
Fig. 1 shows the $m$-distribution at the critical temperature in the $(d,q)=(2,3)$
case, in the first-scaling form. The scaling is recovered very precisely,
even for such small system sizes as $64 \times 64$. 
Note also the complicated shape of the scaling curve.

\begin{figure}[h]
\epsfig{figure=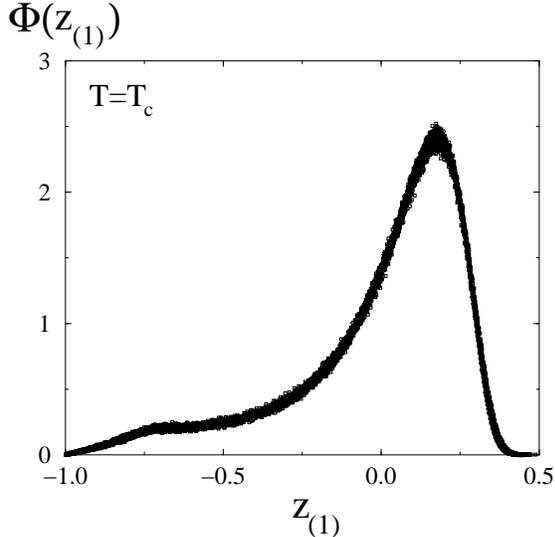,height=7cm}
\caption{Scaled $m$-distribution for the 3-states Potts model ($q=3$) on
2-dimensional $L \times L$ square lattice, at the critical temperature.
Three sizes are shown : $L=32$ (stars), $L=48$ (circles) and $L=64$ (squares).  
The thermalization is achieved after $4\times10^6$ Monte-Carlo steps, starting from the
initial disordered system.}
\label{fig1}
\end{figure}

We can discuss this scaling here in the slightly different context of correlated 
variables \cite{BG}. Consider for simplicity, the Ising model ({\it i.e.} $(d,q)=(d,2)$ case). 
The extensive order parameter is just the sum of $N$ {\it correlated} variables : 
$M=\sum s_i$. 
When the system is disordered, the spins are correlated on the short distance $\xi$ 
($\xi/N \rightarrow 0$ at the thermodynamic limit) and their mean value is zero.
The central limit theorem tells then that the distribution of the random variable 
$M/\sqrt{N}$ is Gaussian when $N$ becomes large, with zero mean and finite 
variance. This can also be expressed by the asymptotic law :
\begin{eqnarray}
<M^2> P[M^2] = \sqrt \frac {<M^2>}{2 \pi M^2} \exp \left( -\frac{M^2}{2<M^2>} \right)
\end{eqnarray}
which is correctly in the first-scaling form, with a Gaussian shape and the 
trivial anomalous exponent $g=1/2$ (see (\ref{ht})).

On the other hand, if the system is in the ordered phase, the average value
of the individual spins is finite, say : $<s_i> = m$, and the same reasoning can
be used for the variable : $(M-Nm)/\sqrt{N}$. This variable is of zero mean ,
finite variance and short-ranged correlated. So, its fluctuations are Gaussian
and can be put in the second-scaling form :
\begin{eqnarray}
<M>^{1/2} P[M] = \frac{1}{\sqrt{2 \pi}} 
\exp \left(-\frac{(M-<M>)^2}{2<M>} \right)  ~ \ .
\end{eqnarray}
Of course, the most interesting case corresponds to the critical temperature. At this
point, the spins are correlated throughout the system, and the magnetization
cannot be evaluated by the central limit theorem. Instead, we can remark that
the spin-correlations are power-law : 
$$<s_{\vec{r}_o} s_{\vec{r}+\vec{r}_o}> \sim \frac {1}{r^{d-2+\eta}}~ ,$$ with 
$\eta$ a critical exponent whose value should be between $2-d$ and 2. Looking
at the total magnetization as the sum of $N$ correlated variables, one gets
\begin{eqnarray}
<M^2> & = & \sum_{i,j} <s_i s_j> = \sum_i <s_i^2> +\nonumber \\
& + & N \int_1^{L} <s_{\vec{0}} s_{\vec{r}}> r^{d-1} dr
\end{eqnarray}
with $L \sim N^{1/d}$ the typical macroscopic length of the system. This means
that : $$<M^2> \sim N^{1+\frac{2-\eta}{d}}$$ as the leading behaviour. 
This non-trivial anomalous exponent $$g=\frac{1}{2} +\frac{2-\eta}{2d}$$ 
between $1/2$ and 1, is here 
the sign of the criticality. The first-scaling law should hold in this case, as for 
the $(d,q)=(2,3)$ Potts model discussed above, but the scaling function should be
different since it depends on the precise form of the interactions. Only the tail 
can be linked to another critical exponent, as it has been written in Sect. II.C.

\section{Reversible aggregation - the percolation model}
Percolation model can be defined as follows. In a box
(a part of the regular lattice), each site corresponds to a monomer and 
a proportion $p$ of active bonds is set randomly between sites (the
bond percolation model).
Such a network results in a distribution of clusters defined as ensemble of 
occupied sites connected by active bonds. For a definite value of $p$, say 
$p_{cr}$, a giant cluster almost surely spans the whole box. 
The sol-gel transition corresponds to the appearance of 'infinite' cluster 
(gel) at a finite time. 'Infinite' in this context means that gel contains 
the finite fraction of total mass of the system. The sol - gel transition in 
finite systems can be suitably studied using moments of the number-size-distribution 
$n_s$, {\it i.e.} the number of finite clusters of size $s$ :
\begin{eqnarray}
\label{eq1}
M_k' = \sum_{s < s_{max}} s^k n_s  ~ \ ,
\end{eqnarray}
where the summation is performed over all clusters with the exception of the 
largest cluster $s \equiv s_{max}$. The superscript $'$~ recalls this 
constraint on summation in Eq. (\ref{eq1}). The mass of the largest cluster 
is then : $N-M_1'$~, with :
\begin{eqnarray}
N = \sum_{{\text all}~ s} s n_s  ~ \ .
\nonumber
\end{eqnarray} 
In infinite systems, one works with the normalized moments of the 
concentration-size-distribution $c_s$, {\it i.e.} the concentration of clusters of 
size $s$ : 
\begin{eqnarray}
\label{eq2}
m_k' = \sum_{s} s^k c_s  ~ \ ,
\end{eqnarray}
where the summation in (\ref{eq2}) runs over all finite clusters. Generally, 
concentrations are normalized  such that :
\begin{eqnarray}
c_s= \lim_{N \rightarrow \infty} \frac{n_s}{N} ~ \ .
\nonumber
\end{eqnarray} 
The probability that a monomer belongs to the infinite cluster (gel) is equal to 
$1-m_1'$~, with : 
\begin{eqnarray}
m_k'= \lim_{N \rightarrow \infty} \frac{M_k'}{N} ~ \ .
\nonumber
\end{eqnarray}

For example, in the thermodynamic limit when the size of the box becomes infinite , a finite 
fraction of the total number of vertices belongs to this cluster. Therefore, we get the 
results : $m_1' = 1$ for $p < p_{cr}$ and $m_1' < 1$ for $p > p_{cr}$. Moreover, $m_1'$ 
is a decreasing function of the occupation probability. This typical behaviour is commonly 
(and incorrectly) called : '{\it the failure of mass conservation}', but, as stated before, 
$m_1'$ is more simply the probability for a vertex to belong to some finite cluster.

\subsection{Percolation on the Bethe lattice}
The bond-percolation on the Bethe lattice with coordination number ${\hat z}$, has been 
solved by Fisher and Essam \cite{Fisher}. Here the main result we are interested in, is 
the concentration-size-distribution \cite{Fisher} :
\begin{eqnarray}
\label{eq4}
c_s ={\hat z}\frac{(({\hat z}-1)s)!}{(({\hat z}-2)s+2)! s!} p^{s-1} 
(1-p)^{({\hat z}-2)s+{\hat z}} \nonumber 
\end{eqnarray}
and the first normalized moment :
\begin{eqnarray}
\label{eq5}
m_1'= \left( \frac {1-p}{1-p^{*}} \right)^{2{\hat z}-2} ~ \ ,  \nonumber 
\end{eqnarray}                          
with $p^{*}$ being the smallest solution of equation : 
\begin{eqnarray}
\label{eq6}
p^{*}(1-p^{*})^{{\hat z}-2} =p(1-p)^{{\hat z}-2} \nonumber ~ \ .
\end{eqnarray}
Let us define : $$p_{cr} \equiv \frac{1}{{\hat z}-1}~ \ .$$ 
For $p < p_{cr}$, the only solution of 
the above equation is : $p^{*} = p$, but when $p$ is larger than $p_{cr}$, then there is 
a smaller non-trivial solution which behaves as $p_{cr}-|p-p_{cr}|$ near $p_{cr}$. Above 
this threshold, the moment $m_1'$ is smaller than 1 and behaves approximately as : 
$$m_1^{'} \simeq 1-\frac{2(p-p_{cr})}{1-p_{cr}}~ \ .$$ The marginal case ${\hat z} = 2$ corresponds 
to the linear-chain case.

Coming back to the concentrations, we can see that for large values of the size $s$, the 
following Stirling approximation holds :
\begin{eqnarray}
\label{eq7}
c_s \sim s^{-5/2} \exp (-\alpha s)     \nonumber 
\end{eqnarray}
\noindent
with $\alpha $ given by :
\begin{eqnarray}
\label{eq8}
\alpha = \ln \left[ \frac {p}{p_{cr}} \left( \frac {1-p}{1-p_{cr}} 
\right)^{{\hat z}-2}  \right] ~ \nonumber \ .
\end{eqnarray}
For this model, a power-law behaviour of the concentrations is seen at the threshold 
$p_{cr}$, namely : $c_s \sim s^{- \tau}$ with $\tau =5/2$. Outside this threshold, 
an exponential cut-off is always present \cite{stau}. 
This sort of critical behaviour at 
equilibrium is analogous to the thermal critical phenomena, and in particular,
there exist two independent critical exponents, for example $\tau $ and $\sigma $. The 
latter one is the exponent of the mean cluster-size divergence. Together, those two
critical exponents : $\tau = 5/2$ and $\sigma =1 $, describe completely the critical 
features. 

This singular behaviour is due to the appearance of a 
giant cluster, the so-called
percolation cluster, at the transition point. More precisely, 
in the infinite system the probability for a given 
site to belong to this infinite 
cluster is zero below the critical threshold $p_{cr}$ and positive above it. This 
probability is non-analytical at the critical point. Because of this behaviour, the 
extensive order parameter defined for finite systems is just the size of the largest 
cluster $s_{max}$. 

As discussed in Sect. II.C, the corresponding finite-size order 
parameter scales as $$s_{max} \sim N^{2/3} ~ \ . $$
Even though the system experiences the second-order critical phenomenon, fluctuations of 
the multiplicity
distribution remain small and the KNO-scaling does not 
hold. Of course, 
$m_0'$ is not in this case an order parameter since $\tau > 2$ even though there is 
some irregularity in its behaviour passing the threshold. This non-analyticity can be 
illustrated by the exact result for bond-percolation on the Bethe lattice. In this 
mean-field case, the normalized $0^{th}$-moment is :
\begin{eqnarray}
\label{eq18}
m_0' & = & (1-\frac {{\hat z}}{2} p^{*}) \left( \frac {1-p}{1-p^{*}} 
\right)^{2{\hat z}-2} \nonumber \\
&\simeq& \frac {{\hat z}-2}{2({\hat z}-1)} -({\hat z}-1) \epsilon + 
(1-\frac {{\hat z}}{2}) |\epsilon |   \nonumber 
\end{eqnarray}
with : $\epsilon = p-p_{cr}$, and $\epsilon \ll 1$. There is a jump of the first 
derivative of $m_0'$ with respect to $p$ : $-{\hat z}/2$ for $p \rightarrow p_{cr}^{-}$,
and $(4-3{\hat z})/2$ for $p \rightarrow p_{cr}^{+}$. 

\subsection{3-dimensional percolation}
As shown by Botet et al \cite{prln}, the multiplicity distribution for the 
3-dimensional bond 
percolation model on the cubic lattice at the infinite-network percolation threshold 
exhibits the $\Delta$-scaling with $\Delta = 1/2$~, and hence the fragment
multiplicity is not related to the order parameter  
in this process. This is shown in Fig. 2c as a typical example
of non-critical parameter scaling. Note that the multiplicity distributions in
Fig. 2c are plotted in semi-logarithmic
form to show clearly the Gaussian behaviour (parabolic shape on the figure).
The proper order parameter for this model is the normalized 
mass of the gel-phase, {\it i.e.} the mass of the largest cluster 
divided by the total mass of the system $s_{max}/N$~. Different probability distributions 
$P_N[s_{max}/N]$~ for different system sizes $N$~ can be all 
compressed into a unique characteristic function (see Fig. 2.a) :
\begin{eqnarray}
\label{eq18a}
<\frac{s_{max}}{N}> P_N\left[\frac{s_{max}}{N}\right] = \Phi \left( \frac{s_{max} - <s_{max}>}{<s_{max}>}
\right) \nonumber
\end{eqnarray}
which is analogous to the KNO-scaling function.  As an application of 
results developed in Sect. II.E, Fig.2b shows the $\Delta$-scaling for the 
shifted order parameter :
$M_1' = N - s_{max}$. The value of $\Delta$ (= 0.8), is consistent with the 
value of the anomalous dimension (25) : $g=0.8435$, for the accepted values of
the critical exponents $\beta, \gamma$ in the 3-dimensional 
percolation \cite{Stauffer}.
One should also remember, that $\Delta$ has been extracted from the
small size ($N=14^3, 20^3, 32^3$) percolation network calculations  
at the {\it infinite-network} percolation threshold. This explains a small
difference between the value for $\Delta$ from the scaling analysis 
and the expected value : $\Delta = g$, in the infinite network.

According to the results derived above for the second order phase transition,
the second-scaling should hold outside of the critical point. This is correctly realized with
the three variables $s_{max}$, $M_1'$ and $M_0$ for large or small values of the probability 
$p$. Fig. 3 shows such results for the value $p=0.35$.
\begin{figure}[]
\epsfig{figure=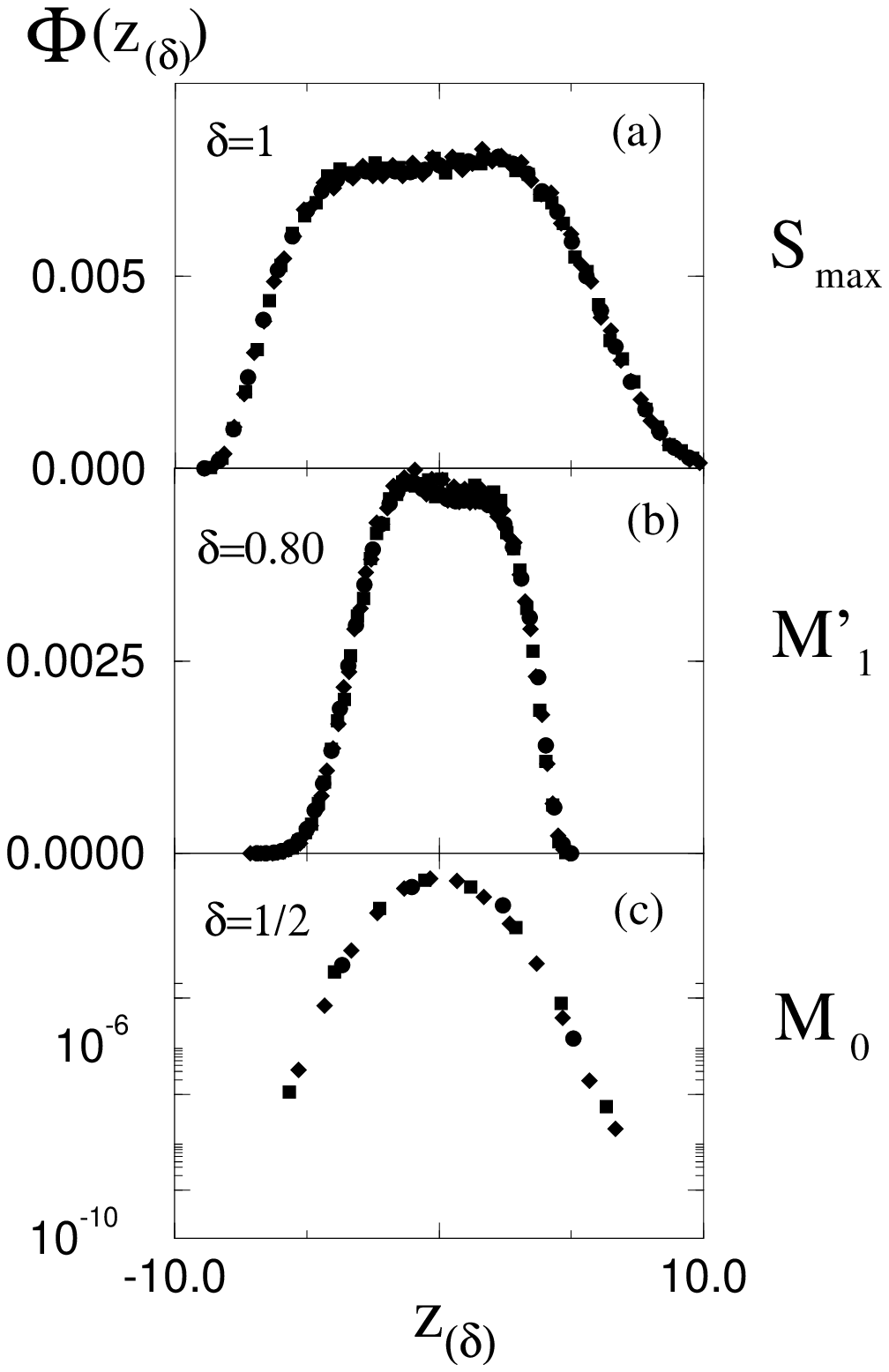,height=13.5cm}
\caption{(a) The first-scaling of $s_{max}$-distributions at the
percolation threshold ($p=p_{cr}=0.2488$) of the 3-dimensional bond percolation
for lattices of different sizes :
$N=14^3$ (diamonds), $N=20^3$ (squares), $N=32^3$ (circles). The data
corresponds to $10^5$ events. \\
         (b) The $\Delta $-scaling of the distributions of 
$M_1' = N-s_{max}$ for the same conditions as in (a). \\
         (c) The second-scaling of the multiplicity distributions
plotted in log-linear scale ({\it i.e.} $\log (<M_0>^{1/2} P[M_0])$ 
vs $z_{(1/2)}$) for the same conditions as in (a).}
\label{fig2}
\end{figure}

Finally, it is instructive to see how the first-scaling is disappearing when
the value of $p$ is slightly shifted away from its critical value. Fig. 4
illustrates the deviations from the first-scaling for the values of 
parameter $p$ close to $p_{cr}$, on both sides of $p_{cr}$.   
Even very close to the critical point, 
these deviations are quite significant and can be easily
seen in this representation.   

\section{Irreversible aggregation - example of Smoluchowski kinetic model}
The irreversible sol-gel transition can be modelled using the coupled non-linear 
differential equations in distributions $c_s$ of clusters of mass $s$ per unit
volume (the Smoluchowski equations \cite{smo}) :
\begin{eqnarray}
\label{eq3}
\frac{dc_s}{dt} = \frac{1}{2} \sum_{i+j=s} K_{i,j} c_i c_j -
\sum_{j} K_{s,j} c_s c_j ~ \ .
\end{eqnarray}
\begin{figure}[]
\epsfig{figure=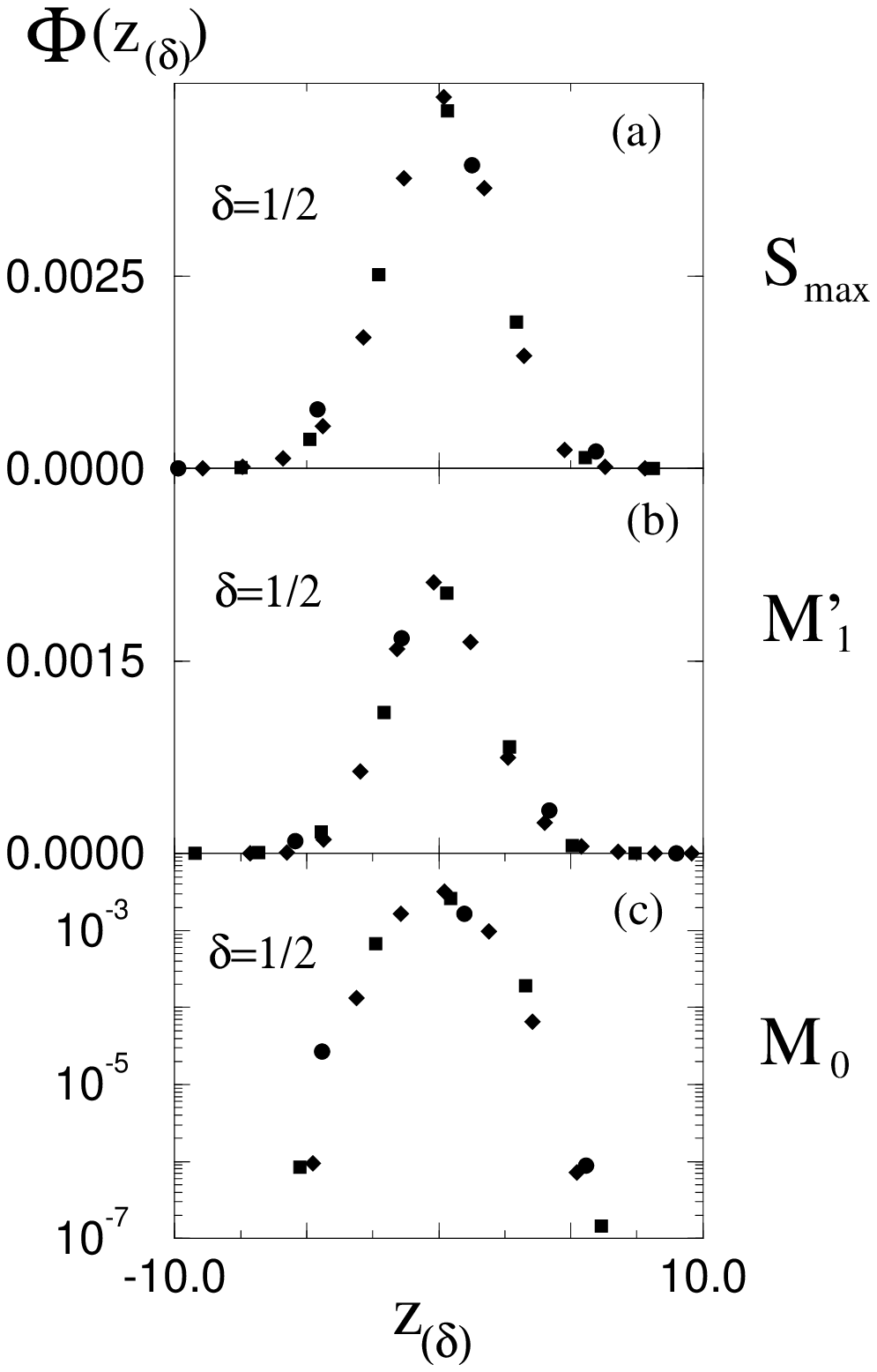,height=14cm}
\caption{(a) The second-scaling of $s_{max}$-distributions above the percolation
threshold ($p=0.35$) of the 3-dimensional bond percolation for lattices of
different sizes :
$N=14^3$ (diamonds), $N=20^3$ (squares), $N=32^3$ (circles). The calculated
data corresponds to $10^5$ events. \\
         (b) The second-scaling of the distributions of 
$M_1' = N-s_{max}$ for the same conditions as in (a).\\
         (c) The second-scaling of the multiplicity distributions
plotted in log-linear scale ({\it i.e.} $\log (<M_0>^{1/2} P[M_0])$ vs $z_{(1/2)}$)
for the same conditions as in (a).}
\label{fig3}
\end{figure}
                                                           
\begin{figure}[]
\epsfig{figure=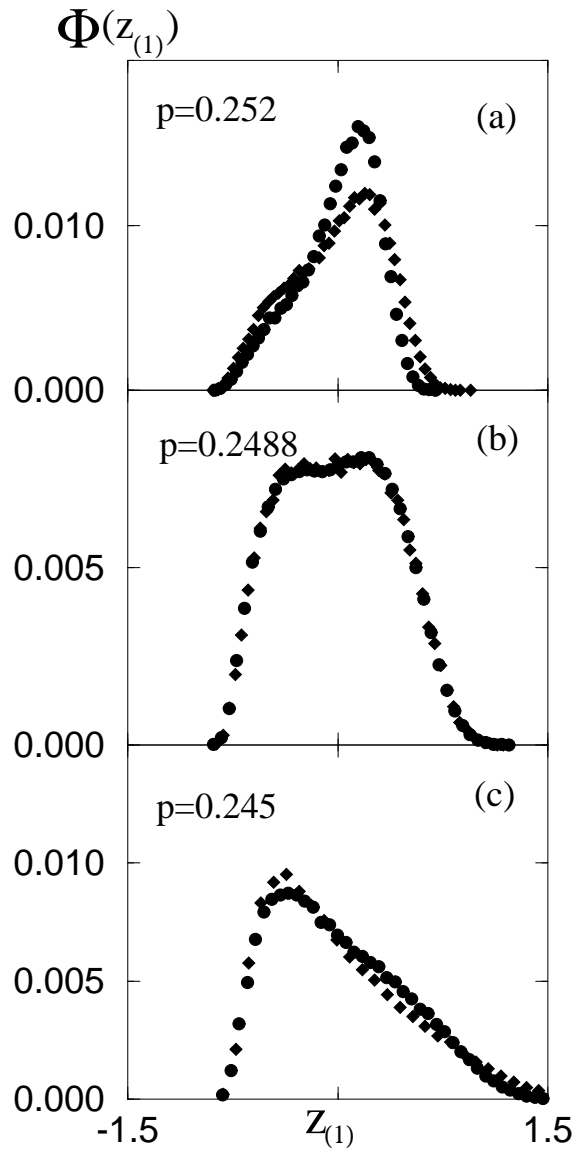,height=15cm}
\vspace{0.7cm}
\caption{(a) The $s_{max}$-distributions are plotted in the first-scaling
form for parameters $p$ close to the percolation threshold of the 3-dimensional
bond percolation : (a) $p=0.252$ ; (b) $p=p_{cr}=0.2488$ and (c) $p=0.245$. 
The calculations are done for lattices of different sizes :
$N=14^3$ (diamonds), $N=32^3$ (circles). The calculated data
corresponds to $10^5$ events.}
\label{fig4}
\end{figure}
Coefficients $K_{i,j}$ represent the probability of aggregation per unit of time 
between two clusters of mass $i$ and $j$ . The Smoluchowski equations are derived from the 
master equation in the mean-field approximation \cite{marcus} : 
$$<c_k c_l> = <c_k><c_l>~ .$$
The time $t$ includes both diffusion and reaction times. 
Eqs. (\ref{eq3}) suppose irreversibility of the aggregation, {\it i.e.} 
the cluster fragmentation is excluded. One should notice however 
that the sum over
$j$ in (\ref{eq3}) does not include the infinite cluster (gel) because : 
$$c_{j = \infty } = 1/\infty = 0~ .$$  

Experimentally known aggregation kernels $K_{ij}$ are homogeneous functions 
\cite{simons} : 
\begin{eqnarray}
K_{ai,aj} = a^\lambda K_{i,j}
\nonumber
\end{eqnarray}
with $\lambda$ being the homogeneity index. Perhaps the simplest physically 
relevant example of a homogeneous kernel is : $K_{i,j}=(ij)^\mu$. It has been 
shown in this case that if $\mu$ is larger than 1/2, then there exists a time 
$t_{cr}$ ($t_{cr} < \infty $) such that $m_1' = 1$ for $t \leq t_{cr}$ but 
$m_1' < 1$ for $t > t_{cr}$  \cite{gel1,gel2}. 

Let us consider now the case : $K_{i,j} = (ij)^{\mu}$ with $\mu = 1$ in more 
details. It was shown in this case \cite{gel1} that the critical gelation time 
is : $t_{cr}=1$, and the solution for size-distribution of the Smoluchowski 
equations with the monodisperse initial condition is \cite{mleod} :
\begin{eqnarray}
\label{eq9}
c_s & = & \frac {s^{s-2}}{s!} e^{-st} t^{s-1} ~~~~~~~  \mbox{ for}  
~~~ t \leq 1  \nonumber \\ \nonumber \\
c_s & = & \frac {s^{s-2}}{s!} \frac{\exp (-s)}{t} ~~~~~~~  \mbox{ for} 
~~~  t > 1  ~~~ \  \nonumber 
\end{eqnarray}
The asymptotic solution for large $s$ is :
\begin{eqnarray}
\label{mono2}
c_s & \sim & \frac{1}{t \sqrt {2\pi}}s^{-5/2} \exp \left[ -s(t-1+\ln t)
\right] ~~~  \mbox{for}  ~~~  t \leq 1   \nonumber \\    \nonumber \\
c_s & \sim & \frac{1}{t \sqrt {2\pi}}s^{-5/2}  ~~~~~~~~~~  \mbox{for}  ~~~~~~  
t > 1  
\end{eqnarray}
Note that the power-law behaviour ($\tau = 5/2$) 
is present for $t > 1$ and not only at the threshold. 
The whole distribution of finite-size clusters evolves self-similarly and the 
appearance of a power-law behaviour is not here a sign of the critical behaviour
but a specific characteristics of the gelation phase.

The solution for the first normalized moment is :
\begin{eqnarray}
\label{eq10}
m_1'& = & 1 ~~~~   \mbox{ for}   ~~~  t \leq 1  \nonumber \\ \nonumber \\
m_1' & = & \frac{1}{t} ~~~  \mbox{ for}  ~~~  t > 1  ~~~ \  \nonumber 
\end{eqnarray}
With those asymptotic forms of $c_s$ one can calculate the gel fraction in the 
infinite system before and after the critical point :
\begin{eqnarray}
\label{mono5}
m_G & = & 0 ~~~~~~~~~  \mbox{for}  ~~  t \leq 1  \nonumber \\  \nonumber \\
m_G & = & 1 - \frac{1}{t} ~~~  \mbox{for}  ~~  t > 1  ~  \nonumber 
\end{eqnarray}

It has been shown that gelation is analogous to a dynamical critical 
phenomenon with \cite{hez} :
\begin{eqnarray}
\label{mono3}
m_G = \lim_{N \rightarrow \infty} \frac{1}{N}<s_{max}> \nonumber 
\end{eqnarray}
as the order parameter. For one realization, $s_{max}$ corresponds to the mass
of the gel above $t_c = 1$. 

For finite sizes, one makes the usual assumption 
that there exists a characteristic size diverging at the transition, say : 
\begin{eqnarray}
N_c \sim |t - 1|^{-1/\sigma_N} 
\end{eqnarray}
such that for the mass gel in the finite system one has :
\begin{eqnarray}
\label{mono6}
\frac{1}{N}<s_{max}>~ \sim (t-1) f\left(\frac{N}{N_c}\right) ~~~~~ 
\mbox{for} ~~~~~ t \geq 1~ \  \nonumber 
\end{eqnarray}
In particular, at the gelation time one has : 
\begin{eqnarray}
\label{mono7}
<s_{max}>~ \sim ~N^{1-\sigma_N} \sim N^g ~ \ .
\end{eqnarray}
Using the formula (\ref{eq28}), which is valid both for the equilibrium and 
non-equilibrium systems, 
one can calculate the anomalous dimension. Given the value of $\tau$ 
(see (\ref{mono2})), one finds $g=2/3$. Hence : 
$\sigma_N = 1/3$ can be deduced from (\ref{mono7}).
The average value of the order parameter $<s_{max}>$ increases 
logarithmically for $t<1$, and is a finite proportion of the 
system size when $t>1$.
\begin{figure}[]
\epsfig{figure=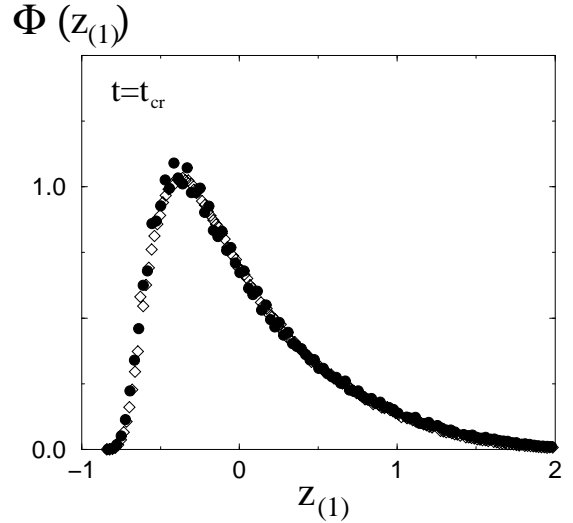,height=7cm}
\caption{The first-scaling of the $s_{max}$ variable in the Smoluchowski 
kinetic model with the kernel $K_{ij}=ij$ at the critical time $t=t_{cr}=1$.
The calculations are performed for two system sizes :
$N=2^{10}$ (diamonds) and $N=2^{14}$ (circles). Each data set corresponds 
to $10^5$ independent events.}
\label{fig5}
\end{figure}

The illustration of the above discussion is shown in Figs. 5 and 6. Fig. 5
shows the distribution of $s_{max}$ in the first-scaling variables for systems
of different sizes. The results have been obtained in the
Smoluchowski model with the kernel : $K_{ij}=ij$, 
at the critical time : $t=t_{cr}=1$. 
Fluctuation properties of $s_{max}$ outside of the critical time : $t=2t_{cr}$,
are shown in Fig. 6. The remaining parameters of the Smoluchowski calculations
are the same as used in the calculations shown in Fig. 5. In this case, 
the data for different system sizes collapse into the
universal curve in the scaling variables with $\Delta = 1/2$. One should keep
in mind that in both cases,
the fragment-size distribution is a power-law with $\tau=5/2$ (see
(\ref{mono2})).

\begin{figure}[]
\epsfig{figure=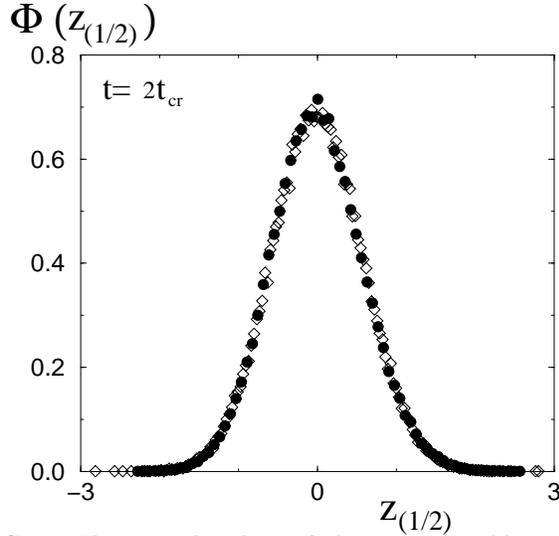,height=7cm}
\caption{The second-scaling of the $s_{max}$ variable in the Smoluchowski 
kinetic model with the kernel $K_{ij}=ij$ above the critical time $t=2t_{cr}=2$.
The calculations are performed for two system sizes :
$N=2^{10}$ (diamonds) and $N=2^{14}$ (circles). Each data set corresponds 
to $10^5$ independent events.}
\label{fig6}
\end{figure}

\begin{figure}[]
\epsfig{figure=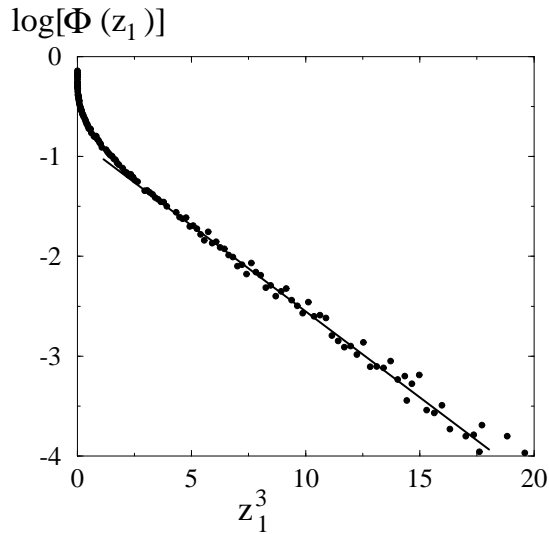,height=7cm}
\caption{The plot of the large - $z_{(1)}$ tail of the decimal logarithm of the
scaling function : $\log \Phi(z_{(1)})$, against $z_{(1)}^3$ for the system 
size $N=2^{12}$. The calculated data correspond to $10^5$ independent events.
The solid line shows the dependence : $\Phi(z_{(1)} \sim
\exp(-z_{(1)}^3)$, which is  expected from the value of the anomalous dimension $g = 2/3$.}
\label{fig7}
\end{figure}

The relation  between the form of tail of the scaling function and the
anomalous dimension (\ref{gdab}) was derived analytically in Sect. II.C for the equilibrium
systems at the second-order phase transition. For non-equilibrium systems, 
we do not know equally rigorous derivation (see also Sect. VI.A). 
On the other hand, one may
expect that the relation between the $N$-dependence of the average value of
the order parameter and the asymptotic form of the scaling function in the
limit $N \rightarrow \infty$, {\it i.e.} between ${\hat \nu}$ and $g$, 
is connected to the asymptotic stability of the limit distributions. Actually,  
there is a very close connection of the renormalization group ideas and the  
limit theorems in the probability theory \cite{bleher}.
If true, then the relation (\ref{gdab}) could be valid in
a more general framework than the one provided 
by the equilibrium statistical mechanics. To check this assertion, in Fig. 7 we
show the plot of the logarithm of the scaling function $\Phi(z_{(1)})$ (see
Fig. 5) versus
$z_{(1)}^3$ for large values of $z_{(1)}$. If the relation (\ref{gdab}) is
valid also for the non-equilibrium sol - gel second-order phase transition,
then : $\Phi(z_{(1)}) \sim \exp(-z^3)$, and the tail of the scaling function
should be a straight line in Fig. 7. That is indeed the case.

Figs. 8 and 9, show the $\Delta$-scaling for the shifted order parameter 
variable : $M_1^{'}=N-s_{max}$. Results of the Smoluchowski calculations
with the kernel : $K_{ij}=ij$, are shown at $t=t_{cr}$ (see Fig. 8),  
and at $t=2t_{cr}$ (see Fig. 9).  One sees that
the $M_1^{'}$ distribution exhibits a qualitative change while going from the 
critical time $t=t_{cr}$, where $\Delta = 0.67$, to $t=2t_{cr}$ for which
$\Delta = 1/2$. At $t=t_{cr}$, the value of $\Delta$ obtained by superposing
different $M_1^{'}$ distributions in the scaling plot (\ref{delta}),
agrees perfectly with the value of the
anomalous dimension $g$ (=2/3).  

\begin{figure}[]
\epsfig{figure=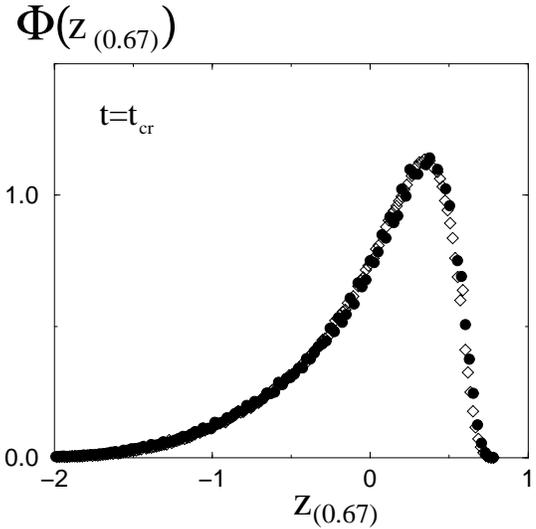,height=7cm}
\caption{The $\Delta$-scaling of the distributions of 
$M'_1$ shifted order parameter in the Smoluchowski kinetic model
with kernel $K_{ij}=ij$ at the critical time $t=t_{cr}=1$.
Two system sizes are considered :
$N=2^{10}$ (diamonds) and $N=2^{14}$ (circles). The
calculated data correspond to $10^5$ independent events.}
\label{fig8}
\end{figure}

By comparing Figs. 5, 6 with Figs. 8, 9, one may see also that
the effect of changing the variable : $s_{max} \longrightarrow M_1^{'}$, is seen
only at the critical time (compare Figs. 5 and 8) where 
$(\Delta = 1) \longrightarrow (\Delta = 0.67)$, and is absent above the critical time (compare Figs. 6 and 9) 
where $\Delta $ (=1/2) remains unchanged.  

\begin{figure}[]
\epsfig{figure=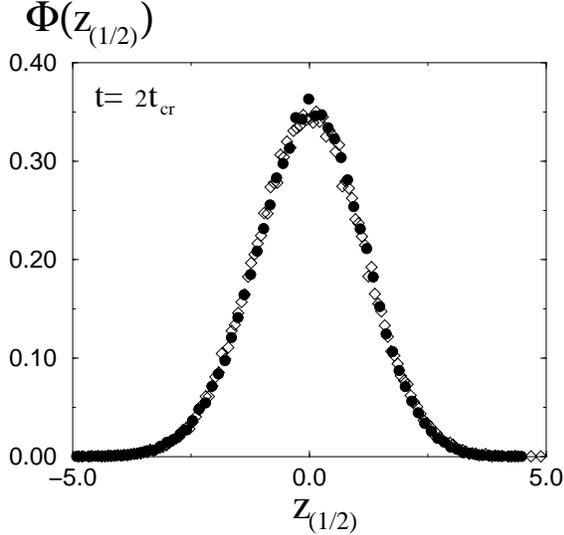,height=7cm}
\caption{The second-scaling of the distributions of 
$M'_1$ shifted order parameter in the Smoluchowski kinetic model
with kernel $K_{ij}=ij$ above the critical time, at $t=2t_{cr}$. 
Two system sizes are considered :
$N=2^{10}$ (diamonds) and $N=2^{14}$ (circles). The
calculated data correspond to $10^5$ independent events.}
\label{fig9}
\end{figure}
\begin{figure}[]
\epsfig{figure=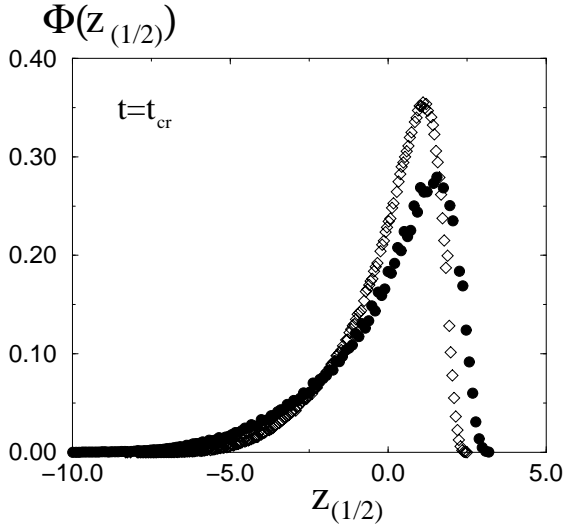,height=7cm}
\caption{The distributions of 
$M'_1$ shifted order parameter in the Smoluchowski kinetic model
with kernel $K_{ij}=ij$ at the critical time $t=t_{cr}=1$ , are plotted in the
scaling variables of second-scaling.
Two system sizes are considered :
$N=2^{10}$ (diamonds) and $N=2^{14}$ (circles). The
calculated data correspond to $10^5$ independent events. The failure of the
second-scaling is clearly visible.}
\label{fig10}
\end{figure}

Finally, in Fig. 10 we show the
size-dependence of the $M_1^{'}$-distributions at $t=t_{cr}$, when the
distributions are plotted in the
'wrong' variables of the second-scaling $\Delta = 1/2$. The distributions for
two system sizes are clearly displaced , 
showing the sensitivity of the scaling analysis and failure of the
second-scaling.

\subsection{The origin of fluctuations and the argument of Van Kampen}
$\Omega$ - expansion is a systematic expansion of the master equations in 
powers of $1/N$ \cite{kampenvan}. Lushnikov was the first to express the 
generating 
functions as the contour integrals for quantities like the moments $M_k'$ 
\cite{lushnikov}. Then Van Dongen and Ernst \cite{vde} 
used $\Omega$ - expansion to 
calculate explicitly these integrals for the moments $M_k'$ 
in some simple cases like :
$K_{ij}=ij$. For example, the result for $M_1'$ can be expressed in terms of 
the generating function for the $s_{max}$-distribution $P_N[s_{max}]$, as :
\begin{eqnarray}
\label{mono12}
\sum_{s_{max}} P_N[s_{max}] e^{s_{max} u} & = & \frac{N!}{2 i \pi} e^{-N u}
\oint \frac{dz}{z^{N+1}} \times \nonumber \\&\times&\exp\left[ \sum_{s=1}^{N} \frac{c_s}{N^{s-1}} 
e^{s(N-s)/(2N)} (z e^{-u})^s \right] \nonumber 
\end{eqnarray}
Using then the identity :
\begin{eqnarray}
\label{monoid}
\frac{\partial^n}{\partial u^n} & &\left[ \exp \left( \sum_s \alpha_s e^{s u}
\right) \right] = 
\sum \frac{n!}{1!^{a_1}a_1!...n!^{a_n}a_n!}\times \nonumber \\ 
&\times&\left[ \sum_s \alpha_s s^1 e^{s u}
\right]^{a_1}...\left[ \sum_s \alpha_s s^n e^{s u} \right] ^{a_n} \exp \left( 
\sum_s \alpha_s e^{s u} \right)  \nonumber 
\end{eqnarray}
, where the sum runs over different sets $\{a_1,...a_n\}$ with the constraint : 
$a_1+...+a_n=n$, and the particular result written down by Van Dongen and Ernst 
for the $K_{ij}=ij$-case \cite{vde} :
\begin{eqnarray}
\label{monoGu}
\exp & &\left[ \sum_{s=1}^{N} \frac{c_s}{N^{s-1}} 
\exp[\frac{s(N-s)}{2N}] z^s \right] =
\nonumber \\ &=&
\sum_{s=1}^{N} \frac{z^s}{s!} \exp[-\frac{1}{2} st\frac{1-s}{N}] + O(1/N) \nonumber 
\end{eqnarray}
we can find :
\begin{eqnarray}
\label{monogs}
\sum_{s_{max}} P_N[s_{max}]&\exp&(s_{max} u) = \nonumber \\&\exp&\left[ N \left( 
\sum_{k=1}^{\infty} 
\frac{m_k' (-u)^k}{k!} + u \right) \right] ~ \ .
\end{eqnarray}
Having the $\Omega$ - expansion of the generating function of the $s_{max}$-distribution, 
we can conclude about the scaling at the gelation point. The moments $m_k'$ 
of the size-distribution for infinite systems are known to diverge 
near the gelation time \cite{Hendriks} as :
\begin{eqnarray}
\label{moments}
m_k' \sim |t-1|^{3-2k} ~ \ . \nonumber 
\end{eqnarray}
For finite systems, using : $\sigma_N = 1/3$, one
obtains :
\begin{eqnarray}
\label{finmom}
m_k' \sim |t-1|^{-2k+3} f_k(N |t-1|^3) \sim N^{\frac{2k-3}{3}}
\end{eqnarray}
at the gelation time. We have then found the asymptotic result :
\begin{eqnarray}
\label{scammk}
N m_k' \simeq a_k <s_{max}>^k ~ \ , \nonumber 
\end{eqnarray}
where $a_k$'s are some positive constants. Inserting (\ref{scammk}) in
(\ref{monogs}), one can show that the 
generating function for $s_{max}$ is the function of a single variable : $<s_{max}> u$,
what is a sufficient condition for the validity of the first-scaling law 
(\ref{iff}).

We can have informations on similar scalings for various moment-distributions. 
$\Omega$ - expansion leads to the results :
\begin{eqnarray}
\label{ommom}
<M_{k}'^2> - <M_{k}'>^2 & = & N m_{2k}' - (1-t) m_{k+1}'^2   \nonumber \\
\nonumber \\
<M_{k}'> & = & N m_k' 
\end{eqnarray}
for the values of $k$ when all the quantities are defined. At 
the transition ($t=1$), the relation 
(\ref{finmom}) allows to calculate the reduced moments $m_k'$. 
The results can be  written in the compact form :
\begin{eqnarray}
\label{scmom}
\frac{<M_{k}'^2>-<M_{k}'>^2}{<M_{k}'>^{2\Delta}} 
\sim {\mbox {const}} ~ \ ,  \nonumber
\end{eqnarray}
with the following values of exponent $\Delta$ :
\begin{eqnarray}
\label{deltas}
\Delta & = & 1/2 ~~~~~~~~~~~~{\mbox {for}}~~~~~k \leq 3/4  \nonumber \\
\Delta & = & 2k/3~~~~~~~~~~{\mbox {for}}~~~~~~3/4 \leq k \leq 3/2  \nonumber \\
\Delta & = & 1~~~~~~~~~~~~~~~{\mbox {for}}~~~~~~3/2 \leq k ~~~ \nonumber  
\end{eqnarray}
These are indications for $\Delta$-scaling according to 
the remark (\ref{scamom}) in Sect. II.H.
More precisely : the moments of order 
$k<3/4$ are not critical (the second-scaling law),
the moments of order $k$ between 3/4 and 3/2 exhibit the $\Delta$-scaling with
$\Delta = 2k/3$. In particular, for $k=1$, one recovers the correct value :
$\Delta = g = 2/3$, corresponding to the general argument of the shifted 
order-parameter (\ref{delta0}) with $a_1 = a_2 = 1$. Finally, when the value of
$k$ is larger than 3/2, we obtain the first-scaling law for the distribution of
moments $M_{k}'$. This is also a consequence of the shifted order parameter
argument, since in these cases : $$<M_{k}'>~ \sim ~<s_{max}^k> ~ \ .$$
Far from the critical point, all the reduced moments $m_k'$ are independent of $N$,
since the correlation size : $$\frac{1}{t-1+\ln t}$$ (see (\ref{mono2})) is finite. Then, for
any value of $k$, the second-scaling law holds, as expected from the general theory.

The above results about $\Delta$-scaling for various moments of the size-distribution
in the Smoluchowski model with kernel $K_{ij}=ij$, are not complete since the arguments
involve only the second cumulant moment $\kappa_2$. 
In principle, as shown in Sect. II.H,
all cumulants should be investigated. So,  
even though many exact results are known
in this model, the complete analytical solution is not yet available.

The same study as presented above for gelling systems, 
can be performed also for 
non-gelling systems. An example of this kind is obtained 
for : $K_{ij} = i+j$. In this case,
the size-distribution is power-law with the exponent $\tau=3/2$
\cite{wyjasnienie} and, following
the discussion in Sect. II.C, the cluster multiplicity can be the 
order parameter. One can derive analytically, that 
the multiplicity distribution is binomial :
\begin{eqnarray}
\label{mono22}
P_N[M_0,t] & = & \left( \begin{array}{c} {N-1}\\{M_0-1} \end{array} \right) 
\left( 1 - \exp (-Nt) \right)^{N-M_0} \times \nonumber \\
&\times&\exp [-(M_0 - 1)Nt] ~ \  \nonumber 
\end{eqnarray}
and can be approximated for $N \rightarrow \infty$ and for a finite value of
$<M_0>/N$ by :
\begin{eqnarray}
\label{mono23}
<M_0>^{1/2}&&P_N[M_0,t] \sim \frac{1}{{\sqrt {2\pi}(1 - e^{-Nt})}}\times
\nonumber \\&\times&\exp \left(
-\frac{1}{2(1 - e^{-Nt})} \frac{(M - <M_0>)^2}{<M_0>} \right) 
\end{eqnarray}
what corresponds to the second-scaling. One may notice, that this binomial
distribution is exactly equivalent to the bond percolation on a Bethe
lattice with the occupation probability : $$p = 1 - \exp (-Nt)~ \ .$$
In spite of self-similar features in the fragment-size distribution at the
infinite time, one does
not see any critical behaviour in the cluster multiplicity distribution at any
time in the non-gelling aggregation systems. This confirms 
the observation made in
Sect. III.A in the Mekjian equilibrium model, that the power-law size-distribution alone
does not guarantee that the system exhibits the critical behaviour.

The insight gained from the numerical simulations of Smoluchowski
equations and the evidences from exact results for both gelling and 
non-gelling aggregation systems, provide strong hints that the
discussion of Sect. II.H is valid not only for the equilibrium systems but also
for the non-equilibrium ones. We see the same significance of the
$\Delta$-scaling in non-equilibrium systems 
as found in thermodynamic systems, and not only at the critical point but also
close to the critical point or even far from it. We believe that this
universality, which is common to equilibrium and non-equilibrium systems, 
has deeper foundation in the relation between renormalization group ideas for
self-similar systems and the limit theorems of probability theory 
for the asymptotic scaling laws of order-parameter distributions. The concept
of statistical equilibrium does not intervene at this level. One should also
remember that the universality discussed in this work, is associated with only 
one critical exponent, and certainly does not exhausts all the singularity 
properties  of the thermodynamical potential in second-order thermal
phase-transition.

\section{Conclusions}
We have presented in this paper the theory of universal scaling laws of the
order parameter fluctuations in any system in which the second-order critical
behaviour exist. These scaling laws, called $\Delta$-scaling laws, 
are rigorously derived for the equilibrium
systems. Moreover, both analytical and numerical evidence is presented 
in favour of the general validity of the $\Delta$-scaling laws also for 
the off-equilibrium processes which exhibit the critical phenomenon of the 
second-order. 

In this work we have discussed different aggregation models, both
reversible and irreversible, finding the same connexion between scaling function
properties and the anomalous dimension (critical exponents). These results can be important in the
phenomenological analysis of 'critical behaviour' in finite systems, 
where the critical exponent analysis is dubious and, moreover, 
the precise mechanism of the process may be unknown. In these cases, the
$\Delta$-scaling analysis allows to select both the relevant
observable and the interesting initial conditions, which
lead to the 'pseudocritical' behaviour in the studied process. Another
interesting aspect of the $\Delta$-scaling analysis is 
the possibility of compressing data and, hence, the elimination of 
redundant dependences in the data on parameters like the system size, the 
total energy, or the total momentum {\it etc.}. This provides an obligatory
intermediate step in any phenomenological analysis before the 
laws governing complicated dynamics can be found. 
Finally, one should stress that
the scaling laws discussed in this paper are independent of 
whether one deals with an equilibrium or an off-equilibrium process. 
This is a crucial advantage in the studies of 
short-lived systems. Examples of
the multifragmentation processes in the collisions  of atomic nuclei 
or atomic clusters well illustrate this problematic
\cite{leshouches}. In the absence of thermal equilibrium, which is a
theoretical hypothesis difficult to verify in dynamically formed
short-lived systems, we simply do not have at our
disposal any other tool to address reliably the question of possible 
'criticality' of the studied process. 

As said before, the $\Delta$-scaling analysis developed in this work provides an
alternative to the critical exponents analysis in the equilibrium systems
and is the only tool for the analysis of the non-equilibrium systems. All
essential information can be deduced from the scaling function, the value of
$\Delta$ parameter, the form of the tail of the scaling function and the value
of the anomalous exponent. With these informations it is possible to find out 
whether the studied process is at the critical point, in its
neighbourhood or far away from it. Reference point in this analysis is the
(approximate) self-similarity of the system. Generalization of the above
scaling theory to
discontinuous phase-order transitions for which the characteristic length can
be defined, is in progress.

\vspace{0.7cm}
\noindent
We thank  R. Paredes V. for providing us with the Potts model data 
which were used in preparing Fig. 1.


\begin{thebibliography}{99}
\bibitem{comment2}
In general, the second scaling law (\ref{second}) can be satisfied in 
a broader class of scaling functions than Gaussian functions\cite{prln}~.
\end{thebibliography}
\end{document}